\documentclass[sigconf,screen]{acmart} %manuscript,screen,review, anonymous
\usepackage{subcaption}
\usepackage{colortbl}
\usepackage{array}
\usepackage{ragged2e}
\AtBeginDocument{%
  }

\copyrightyear{2025}
\acmYear{2025}
\setcopyright{cc}
\setcctype{by}
\acmConference[UIST '25]{The 38th Annual ACM Symposium on User Interface Software and Technology}{September 28-October 1, 2025}{Busan, Republic of Korea}
\acmBooktitle{The 38th Annual ACM Symposium on User Interface Software and Technology (UIST '25), September 28-October 1, 2025, Busan, Republic of
Korea}\acmDOI{10.1145/3746059.3747704}
\acmISBN{979-8-4007-2037-6/2025/09}

\usepackage{pifont}
\usepackage{xspace}
    \usepackage{tikz}
    \usepackage{graphics}
    \newcommand*\frontaleye{%
       \scalebox{0.25}{
\tikzset{every picture/.style={line width=0.75pt}}   
\begin{tikzpicture}[x=0.75pt,y=0.75pt,yscale=-1,xscale=1]
\draw  [fill={rgb, 255:red, 0; green, 0; blue, 0 }  ,fill opacity=1 ] (84.83,179.7) .. controls (84.92,169.21) and (100.66,160.85) .. (119.99,161.01) .. controls (139.32,161.18) and (154.92,169.81) .. (154.83,180.3) .. controls (152.72,170.91) and (137.95,163.54) .. (119.97,163.39) .. controls (101.99,163.23) and (87.1,170.35) .. (84.83,179.7) -- cycle ;
\draw  [fill={rgb, 255:red, 0; green, 0; blue, 0 }  ,fill opacity=1 ] (102.39,180.93) .. controls (102.39,171.24) and (110.25,163.39) .. (119.93,163.39) .. controls (129.62,163.39) and (137.47,171.24) .. (137.47,180.93) .. controls (137.47,190.61) and (129.62,198.47) .. (119.93,198.47) .. controls (110.25,198.47) and (102.39,190.61) .. (102.39,180.93) -- cycle ;
\draw  [draw opacity=0][fill={rgb, 255:red, 255; green, 255; blue, 255 }  ,fill opacity=1 ] (107,174) .. controls (107,171.24) and (109.24,169) .. (112,169) .. controls (114.76,169) and (117,171.24) .. (117,174) .. controls (117,176.76) and (114.76,179) .. (112,179) .. controls (109.24,179) and (107,176.76) .. (107,174) -- cycle ;
\draw  [fill={rgb, 255:red, 0; green, 0; blue, 0 }  ,fill opacity=1 ] (154.65,179.7) .. controls (154.65,190.58) and (139.02,199.41) .. (119.74,199.41) .. controls (100.46,199.41) and (84.83,190.58) .. (84.83,179.7) .. controls (85.71,190.15) and (101,198.47) .. (119.74,198.47) .. controls (138.48,198.47) and (153.77,190.15) .. (154.65,179.7) -- cycle ;
\end{tikzpicture}}
\,}

\usepackage{tcolorbox}
\usepackage{hyperref}

\MakeRobust{\ref}

\makeatletter
\newcommand{\labeltext}[2]{%
  \@bsphack
  \csname phantomsection\endcsname % in case hyperref is used
  \def\@currentlabel{#1}{\label{#2}}%
  \@esphack
}
\makeatother

\newtcbox{\highlight}[1][magenta]{on line, arc=0pt,colback=#1!10!white,colframe=#1!50!black, before upper={\rule[-3pt]{0pt}{10pt}},boxrule=1pt, boxsep=0pt,left=4pt,right=3pt,top=2pt,bottom=1pt}

\newcommand{\researchq}[2]{
    \noindent % No indent please :)
    \highlight{\textbf{\texttt{\textcolor{gray}{#1}}}} % Creates the coloured box for the label
    \textit{#2} % Displays the actual text of the question in italics, keeping all additional formatting
    \labeltext{\highlight{\textbf{\texttt{\textcolor{gray}{#1}}}} \textit{#2}} % "Saving" the rq incl. question text. It's placed invisibly at the position of the actual question, so it can be easily referenced and linked to.
    {rq:#1} % The specific label for this rq; label-name is equivalent to "rq:" + arg1
}

\newcommand{\notextrefrq}[1]{\hyperref[rq:#1]{\highlight{\textbf{\texttt{\textcolor{gray}{#1}}}}}}

\newcommand{\VIPSIM}{\textsc{VIP-Sim}\xspace}
\begin{document}

\title{VIP-Sim: A User-Centered Approach to Vision Impairment Simulation for Accessible Design}

\author{Max R{\"a}dler}
\email{max.raedler@uni-ulm.de}
\orcid{0000-0002-5413-2637}
\affiliation{%
  \institution{Ulm University}
  \city{Ulm}
  \country{Germany}
}

\author{Mark Colley}
\email{m.colley@ucl.ac.uk}
\orcid{0000-0001-5207-5029}
\affiliation{%
  \institution{UCL Interaction Centre}
  \city{London}
  \country{United Kingdom}
}

\author{Enrico Rukzio}
\email{enrico.rukzio@uni-ulm.de}
\orcid{0000-0002-4213-2226}
\affiliation{%
  \institution{Ulm University}
  \city{Ulm}
  \country{Germany}
}

\renewcommand{\shortauthors}{R{\"a}dler et al.}

\definecolor{lightorange}{RGB}{255, 230, 204}
\definecolor{orange}{RGB}{255, 153, 51}

\begin{abstract}
People with vision impairments (VIPs) often rely on their remaining vision when interacting with user interfaces. Simulating visual impairments is an effective tool for designers, fostering awareness of the challenges faced by VIPs.
While previous research has introduced various vision impairment simulators, none have yet been developed with the direct involvement of VIPs or thoroughly evaluated from their perspective.
To address this gap, we developed VIP-Sim. This symptom-based vision simulator was created through a participatory design process tailored explicitly for this purpose, involving N=7 VIPs. 21 symptoms, like field loss or light sensitivity, can be overlaid on desktop design tools. Most participants felt VIP-Sim could replicate their symptoms.
VIP-Sim was received positively, but concerns about exclusion in design and comprehensiveness of the simulation remain, mainly whether it represents the experiences of other VIPs.
\end{abstract}



\begin{teaserfigure}
  \includegraphics[width=\textwidth]{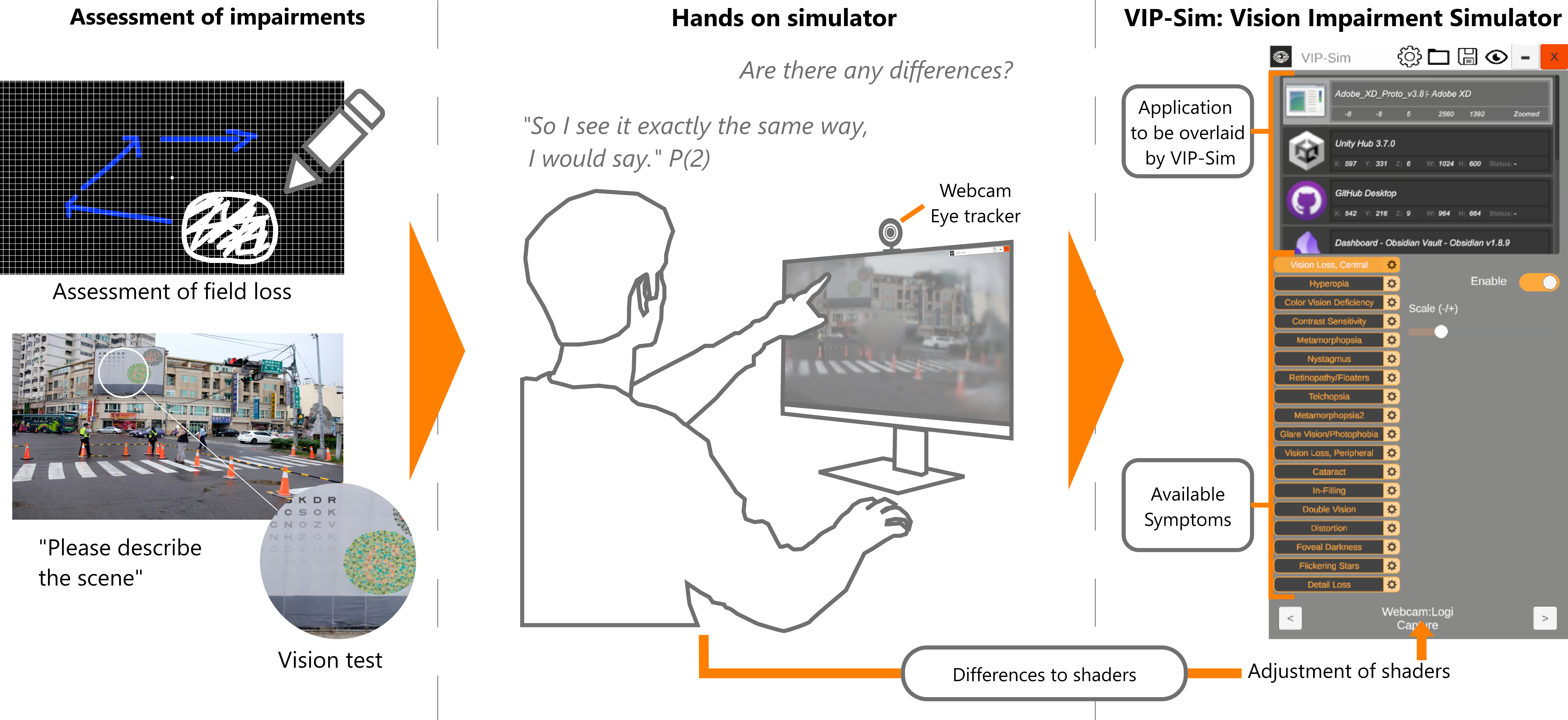}
  \caption{\VIPSIM, a vision impairment simulator, was created in a user-centered design process in collaboration with participants with visual impairments. During this process, \VIPSIM was first built upon existing work and then extended and optimized towards a closer representation of the perception of our participants. This was done in a process where each participant underwent an initial assessment (left) to identify blind spots in their visual fields and descriptions of their perception, combined with vision tests. Subsequently, participants were shown shaders designed to simulate visual impairments (middle). Discrepancies between the participants' actual perceptions and the shader simulations were identified. These differences were either integrated into the existing shaders or used to develop new ones. This process led to the creation of \VIPSIM (right), a symptom-based simulator that can be overlaid on \textbf{any} design tool.}
  \Description{This figure illustrates the user-centered design process. The figure is divided into three sections, arranged from left to right, connected by arrows in that direction. The first section on the far left is labeled "Assessment of Impairments." Under this label is an image of a grid with drawings and the caption "Assessment of Field Loss." A second image in this section shows a city scene, which includes vision tests for color vision deficiencies and contrast sensitivity on a billboard. Below the image, a sentence in quotation marks reads, "Please describe the scene". The middle section is labeled "Hands-On Simulator" and shows a rotoscoped image of a man pointing at a screen. The screen displays a visually altered image, representing a simulation. Two comments are shown in quotation marks: "Are there any differences?" and "So I see exactly the same way, I would say" (P2). Additionally, a webcam is attached to the computer screen, labeled "Webcam Eye Tracker". The right section of the figure displays the user interface (UI) of the simulator, labeled "VIP-Sim: Vision Impairment Simulator." This section contains two descriptions: one for the top part of the UI, where software like AdobeXD can be selected, with the caption "Application to be overlaid by VIP-Sim." The other description points to a list of symptoms, such as "Field Loss Central," labeled as "Available Symptoms". Between the "Hands-On Simulator" and "VIP-Sim: Vision Impairment Simulator" sections, there is an additional arrow labeled "Differences to Shaders and Adjustment of Shader", connecting the image of the man to the UI.}
  \label{fig:teaser}
\end{teaserfigure}

%\received{13 September 2024}

%%
\maketitle

\section{Introduction}
Visual impairment affects over 2.2 billion people worldwide~\cite{stats2023who}. While refractive errors often can be corrected with glasses, conditions such as macular degeneration (8 million) and cataracts (94 million) are irreversible~\cite{stats2023who}. Recent studies highlighted a lack of inclusion of people with vision impairment (VIPs) who can utilize their remaining vision in the designs of visualizations~\cite{kim2021Accessible, wang2024How}. VIPs use their remaining vision and often rely on magnification glasses or screen readers to compensate for their impairments when interacting with interfaces on, for example, displays~\cite{islam2023spacex} or navigating public spaces~\cite{rai2019Coping}. Software designers face the challenge of creating accessible designs, often adhering to guidelines such as the Web Content Accessibility Guidelines (WCAG)~\cite{wcag2024} for legal compliance. These guidelines, while useful, are too general and lack empathy~\cite{kim2018EmpathD, power2012Guidelines, newell2000User}. A vision impairment simulator can be a complementary feature for understanding the challenges VIPs face for designers to also better fulfill guidelines~\cite{choo2019Examining}.

Vision impairment simulators are tools to visualize the effects of visual impairments for sighted users by modifying their perception. %This is done using, for example, spectacles that occlude parts of the vision~\cite{goodman-deane2007Equipping}. 
These simulators offer one approach to communicate difficulties faced by VIPs in daily life~\cite{jones2020Seeing}. Previous research has beneficially employed vision impairment simulators to investigate their impact on designers and their designs~\cite{choo2019Examining, cornish2014Designer, krosl2023Exploring}. Furthermore, vision impairments and their symptoms are complex, and recruiting participants with visual impairments for studies can be challenging~\cite{sears2011Representing}. As VIP feedback is highly valuable, a simulator might help uncover minor problems before testing any design with VIPs, so that the testing with VIPs can focus on more critical issues. It is crucial to acknowledge that simulators cannot fully replicate the experience, the associated challenges of living with a disability~\cite{nario-redmond2017Crip, bennett2019Promise}, or the coping strategies learned and employed.

Almost no studies evaluated their simulated impairments with VIPs~\cite{kasowski2023Systematica}. Works such as \textit{OpenVisSim} by \citet{jones2020Seeing} involved only a single person in the process. \textit{CatARact}~\cite{krosl2020CatARact} focused on only one impairment (cataracts). The exclusion of VIPs in designing such simulators almost certainly leads to inaccurate representations. For example, depicting age-related macular degeneration (AMD) as a black blob occluding parts of the visual field is described as unrealistic and negatively received by those affected~\cite{taylor2018Seeing}. In addition, VIPs themselves have expressed the desire and emphasized the importance of involvement in development to prevent misinterpretation~\cite{leo2016Simulating}. 

Besides needing to involve VIP in the design process of such simulators, many require additional hardware, such as spectacles~\cite{goodman-deane2007Equipping}. Gaze-contingent visualizations, crucial to accurately represent visual impairments, require additional hardware like an eye tracker to mirror the dynamic nature of many visual impairments in real-world scenarios. However, designers want easily accessible tools without additional effort~\cite{tigwell2021Nuanced}.

To address these challenges, we present \VIPSIM, an overlay software impairment simulator that can be applied to the screen of any Windows or macOS machine developed in a participatory design process with N=7 VIPs. \VIPSIM allows the underlying design tool, such as Adobe XD or Unreal Engine, to remain entirely usable while requiring only a webcam for eye tracking.

We developed and conducted a user-centered design (UCD) process with N=7 VIPs to develop visual effects that resemble the affected visual perception of 21 symptoms with adjustable severity (see \autoref{table:shaders}). During this process, \VIPSIM was initially built on existing work and then expanded and optimized to better reflect the perceptions of our participants. We repeatedly applied a qualitative methodological approach (called \textit{MonoWitness Protocol}) in which we first assessed the participants' perception using visual acuity tests (i.e., the Amsler Grid~\cite{amsler_grid_wikipedia}, Ishihara test plates~\cite{ishihara1951tests}, and a contrast sensitivity test~\cite{pelliDesigning1987}; see \autoref{fig:teaser}). These tests helped describe the participants' impairments and identify suitable shaders. The VIPs then compared the simulated impairments with their perception. The approach also involved checking whether the participants' symptoms could be replicated and identifying any necessary adjustments. After each iteration, modifications (shaders) were added to \VIPSIM to reflect the participants' perceptions more closely. A semi-structured interview was also conducted to gather their perspectives on \VIPSIM. Most participants (n=5) felt that \VIPSIM effectively replicated their visual impairment and was well-received by the majority (n=4). While participants expressed only slight concerns about being simulated when they felt \VIPSIM represented their impairment, fear of exclusion, and comprehensiveness, especially regarding the experiences of other VIPs, remain.

\textit{Contribution Statement:}
The main contributions of our work are:

\begin{enumerate}
    \item \VIPSIM, an open-source symptom-focused visual impairment simulator that utilizes webcam-based eye tracking to overlay any Windows or macOS desktop design tool as an artifact contribution. \VIPSIM was developed through an iterative UCD process involving N=7 VIPs to represent their perceptual experiences.
    
    \item The \textit{MonoWitness Protocol}: A detailed description and recommendations for a participatory UCD process and assessment methodology for future work.
    
    \item A qualitative evaluation of \VIPSIM and an exploration of VIPs' perspectives regarding the use of impairment simulators by designers, extending previous work in this area. This evaluation emphasizes the importance of careful and responsible use of impairment simulators.
\end{enumerate}

\section{Related Work}
\label{sec:rw}
This section reviews the field, modalities, and methods of visual impairment simulation. We discuss their capabilities, limitations, and the challenges identified.

\subsection{Visual Impairment Simulation}
\label{sec:rw_vissim}
Visual impairment simulation can promote understanding and empathy for the challenges faced by VIPs~\cite{macnamara2021Simulating}. %Scientific research is also a field that makes use of simulators~\cite{abraham2024Simulation} regarding performance tasks~\cite{juniat2019Understanding} or the impact of visual impairments~\cite{lee2020Effects, wood2010Effect}.  
Various technologies, including wearable simulators (e.g., glasses) and software applications, allow designers and professionals to experience the effects of visual impairments such as macula degeneration, cataracts, and glaucoma~\cite{abraham2024Simulation}. These simulators can create visual effects that alter the perception of objects and environments by, for example, blurring parts of them, allowing users to understand VIPs' difficulties. Studies have shown that such simulations not only increase awareness of barriers and foster empathy but also might help to develop more inclusive designs that meet the needs of VIPs~\cite{pratte2021Evoking, krosl2023Exploring, choo2019Examining}.

As VIP feedback is highly valuable, a simulator might help identify minor issues in advance, allowing subsequent testing with VIPs to focus on more significant problems. Simulators could be particularly useful in addressing areas where current guidelines may fall short, such as fostering empathy and ensuring the completeness of design solutions~\cite{kim2018EmpathD}. Unfortunately, user trials are often time-consuming and costly; simulations like \VIPSIM might offer some benefit in alleviating these challenges by providing a low-cost alternative, though it \textbf{cannot} and \textbf{should not} entirely replace the need for comprehensive user testing.

\subsubsection{Wearable Hardware-based Simulators}
\label{sec:rw_wear}
Simulation spectacles typically use modified lenses to obscure vision with beneficial effects. \citet{zagar2010Low} developed static low-cost glasses by applying materials like fingernail polish or markers to investigate the perceived severity of disease-specific characteristics with sighted participants~\cite{zagar2010Low}. % Participants assessed the severity of the impairment and completed the task.
The Cambridge Simulation Glasses simulate decreasing visual acuity by reducing the perceivable detail~\cite{anderson2013Contemporary, goodman-deane2014Simple, inclusiveDesignToolkit}. By using multiple layers of glasses, different severities can be simulated. However, there is a lack of immersive field loss or light sensitivity due to the lack of eye tracking and technical feasibility. 
Another pair of glasses is the "fork in the Road Vision Rehabilitation LLC Services Goggles", simulating both visual acuity loss and visual field defects of 5 different vision impairments (diabetic retinopathy, glaucoma, cataracts, macular degeneration, and hemianopsia)~\cite{lowVisionSimulators}. \citet{peters2018Effects} used them to evaluate the performance of a brain-computer interface with sighted participants that requires looking at symbols to type. They found that low visual acuity simulation had no effect, while performance decreased with ocular motility impairments. \citet{rousek2011Use} let sighted participants use the fork goggles to reveal wayfinding difficulties, such as inaccessible signage, successfully. However, glasses are not gaze-contingent. In reality, visual impairments like field losses move with the gaze, and for accurate impairment simulation, this is a crucial factor. \citet{zhang2022Seeing} presented a more dynamic set of glasses that leverages two liquid crystal display panels to occlude (darken) parts of the user's gaze, resulting in increased awareness and empathy. Another option to include gaze contingency is contact lenses~\cite{almutleb2018Simulation, macnamara2021Simulating, franceswalonker1981Simulating}. 

The wearable simulators limit themselves to a very small number of symptoms and capabilities. 
In addition, visual impairments like retinitis pigmentosa, glaucoma, or AMD affect parts of the eye where gaze dependency is indispensable. Also, AMD causes visual distortion or bending of straight lines. Spectacles do not cover these complex visual effects and require a more computational approach. With \VIPSIM, we implemented a system capable of simulating diverse, \textbf{customizable symptoms to encompass a wide range of impairments} that incorporate gaze-contingent visual effects.

\subsubsection{On-Desktop Simulators}
\label{sec:rw_desktop}
Existing related work enables on-screen visual impairment simulation. %, where visual impairments are visible on screens.
\citet{kokate2022Exploring} presented an overview of commercial third-party plugins for various design software like Figma, Sketch, or Adobe XD. These plugins aim to visualize and, therefore, uncover issues related to color blindness, contrast deficiencies, and low acuity.

Regarding research, Kamikubo et al. investigated the effects of simulated tunnel vision attached to the user's gaze on a monitor by coloring the screen white offside the tunnel. The study suggests that the simulation-based approach enables developers to observe interface issues and engage with user feedback from participants using the simulator~\cite{kamikubo2018Exploring}.
Goodman-Deane et al. simulated the effect on images. Various visual impairments, such as AMD, can be simulated by selectively occluding portions of the visual field~\cite{goodman-deane2007Equipping}. However, their simulator lacks eye-tracking capabilities and can only be applied to static images. Our \VIPSIM can also apply to animated content like video material.
The DIAS system by \citet{giakoumis2014Enabling} allows user interface (UI) developers to simulate 13 vision impairments as an overlay for Java applications. The blind spots that result from impairments follow the mouse pointer~\cite{giakoumis2014Enabling}, which fails to represent the behavior of these impairments as they are gaze-contingent. \VIPSIM provides gaze-contingent visualizations while applying to any application.

%Mankoff et al. presented the Evaluating Accessibility through Simulation of User Experience (EASE) tool, which also allows for screen-based application of impairment filters~\cite{mankoff2005Evaluatinga}.
%Schulz et al. published a framework for visual deficiency simulation by applying visual effects like blurriness or occlusion of objects~\cite{schulz2019Frameworka}. 
%Furthermore, works such as those by \citet{krishnan2019Impact} or \citet{lane2019Caricaturing} utilize impairments applied to the screen to address quantitative questions like the performance on reading speed. These types of work typically do not involve a tool feasible for designers as it is specially crafted for study purposes~\cite{macnamara2021Simulating}.
%Taylor et al. investigated how participants with AMD see and perceive the world~\cite{taylor2018Seeing}. In their work, the visual effects that VIPs perceive go beyond coloring parts of the visual field black, where fuzziness, sparkles, or bending are described, which are not apparent in existing simulators.

Mankoff et al. introduced the EASE tool, which applies screen-based impairment filters~\cite{mankoff2005Evaluatinga}. Schulz et al. created a framework to simulate visual impairments like blurriness and object occlusion~\cite{schulz2019Frameworka}. \citet{krishnan2019Impact} and \citet{lane2019Caricaturing} used similar screen-based impairments to measure effects on tasks like reading speed, but these are mostly for research, not practical design tools~\cite{macnamara2021Simulating}. The use of eye tracking in desktop applications was also investigated in previous work. Similar to applications of foveated rendering, where only a small region at the center of visual attention is rendered in high detail~\cite{baudisch2003Focusing}, prior research has also utilized this technique to simulate gaze-contingent glaucomas on screens \cite{perry2002Gazecontingent}. Taylor et al. studied visual perceptions of people with AMD, describing distortions like fuzziness, sparkles, or bending, which current simulators rarely show~\cite{taylor2018Seeing}.

Therefore, to develop \VIPSIM, we prompted users about the appropriateness of our simulated symptoms. Another issue with this simulation software is its static nature. They apply to an image or single screenshot, which prevents the perception of animated content when simulating impairments.

\subsubsection{Virtual Reality - Augmented Reality and Smartphone Simulators}
\label{sec_rw_vrar}
Mixed Reality (virtual reality (VR) and augmented reality (AR)) offers immersion by displaying dynamic content projected into the user's eyes. However, mixed reality simulators depend on additional hardware that might not be available and can be cumbersome while working with design tools. Designers desire more straightforward access to such tools~\cite{tigwell2021Nuanced}.

Krösl et al. investigated the effects of cataract simulations in AR and presented a simulator that applies visual effects on a 360-degree video. The prototype uses eye-tracking, and the visual effects were evaluated by participants affected by cataracts~\cite{krosl2020CatARact}. Later, Krösl et al. investigated how an AR simulator (XREye) would influence sighted users and found a positive effect of XREye on sympathy and empathy towards VIPs. They state that the visual effects are based on reports of VIPs~\cite{krosl2023Exploring}.

The OpenVisSim system by Jones et al. presents a VR/AR approach to glaucomatous visual field loss. They also simulated other visual effects like blurriness or glare. These effects were generated based on medical publications and the report of a single patient. Their results regarding task performance with sighted participants indicate that the simulator led to a similar experience for sighted participants as impaired ones~\cite{jones2020Seeing}.

%Ates et al. compared VIZSIM, a head-mounted simulator, with a smartphone-based simulator. In their study, participants were looking at an inaccessible interface. Their results indicate that a head-mounted simulator benefits immersion. However, their tools were not developed based on the recommendations of visually impaired users. Also, it does not use eye-tracking~\cite{ates2015Immersive}. 

Furthermore, Chow-Wing-Bom investigated how degraded visual fields contribute essential information for performing everyday tasks visually-guided using the OpenVisSim simulator~\cite{chow-wing-bom2020Worse}. Yao et al. investigated the interaction with public screens and the challenges encountered using a VR simulator without eye tracking~\cite{yao2021Evaluating}. However, they also state that the visualizations of impairments might not be realistic~\cite{yao2021Evaluating}. Häkkilä et al. developed a VR simulator to investigate the effect on design education and found a positive impact on empathy~\cite{hakkila2018Introducing}. Zhang and Codinhoto report a VR simulator and its positive impact on architects' design. However, here also, eye-tracking is not applied, and no VIPs were involved~\cite{zhang2020Developing}.

\citet{taewoo2024watchcap} used a simulation of field loss to evaluate an optimization device for their visual scanning behavior, which was later assessed with VIPs. Their results from simulated users were successfully used to have a more generalizable evaluation of their system.

Lastly, Kim et al. presented Empath-D, a smartphone interaction simulator. Empath-D visualizes visual impairments in extended reality (XR). The users see a virtual phone in their hand while viewing a passthrough of their arm. At the same time, they are physically holding a turned-off proxy phone. Although the simulator allows for customization of severity, there was no use of eye-tracking and no feedback from VIPs~\cite{kim2018EmpathD}.

\begin{small}
\begin{table*}[ht]
    \centering
    \scriptsize
    \caption{Overview of the simulators mentioned in \autoref{sec:rw}. "n.a." indicates that no information is provided.}
    \begin{tabular}{|p{3cm}|p{2.5cm}|p{1cm}|p{1cm}|p{2cm}|p{3cm}|l|}
    \toprule
    \hline
        Publication & Material & Number of Impairments & Eyetracking & Involvment of VIPs & Developed in UCD\\ \hline
        \citet{zagar2010Low} & Special Glasses & 5 & no & no & no \\ \hline
        \citet{goodman-deane2014Simple} & Special Glasses & 1 & no & \textbf{yes} & Unspecified\\ \hline
        \citet{lowVisionSimulators} & Special Glasses & 5 & no & n.a. & no\\ \hline
        \citet{zhang2022Seeing} & Special Glasses & 3 & yes & no & no \\ \hline
        \citet{almutleb2018Simulation} & Special Contacts & 1 & yes & no & no \\ \hline
        \citet{franceswalonker1981Simulating} & Special Contacts & 3 & yes & no & no \\ \hline
        %--Desktop-- & ~ & ~ & ~ & ~ & ~ & ~ \\ \hline
        \citet{kamikubo2018Exploring} & Desktop PC (Screenshots) & 1 & yes & no & no \\ \hline
        \citet{goodman-deane2007Equipping} & Desktop PC (Screenshots) & n.a. & no & no & no \\ \hline
        \citet{giakoumis2014Enabling} & \textbf{Desktop PC (Applications)} & 8 & no & \textbf{yes} & Unspecified \\ \hline
        \citet{mankoff2005Evaluatinga} & \textbf{Desktop PC (Applications)} & 3 & no & no. & no \\ \hline
        \citet{schulz2019Frameworka} & Desktop PC (Screenshots) & 5 & no & no & no \\ \hline
        \citet{krishnan2019Impact} & Desktop PC (Screenshots) & 1 & no & no & no \\ \hline
        \citet{lane2019Caricaturing} & Desktop PC (Screenshots) & 1 & no & no & no \\ \hline
        \citet{perry2002Gazecontingent} & Desktop PC (Video) & 2 & no & no & no \\ \hline
        %--VR/AR-- & ~ & ~ & ~ & ~ & ~ & ~ \\ \hline
        \citet{krosl2020CatARact} & AR Headset & 1 & yes & \textbf{yes} & Assessed Descriptions \\ \hline
        \citet{krosl2023Exploring} & AR Headset & 4 & yes & no & no \\ \hline
        \citet{jones2020Seeing} & AR Headset & 1 & yes & \textbf{yes} & Uspecified with 1 Participant\\ \hline
        \citet{jones2018Degraded} & AR Headset & 5 & yes & n.a. & no \\ \hline
        \citet{ates2015Immersive} & AR Headset & 6 & no & no & no \\ \hline
        \citet{yao2021Evaluating} & VR Headset & 3 & no & no & no \\ \hline
        \citet{hakkila2018Introducing} & VR Headset & 4 & no & no & no \\ \hline
        \citet{zhang2020Developing} & VR Headset & 4 & no & no & no \\ \hline
        \citet{taewoo2024watchcap} & VR Headset & 1 & yes & no & no\\ \hline
        \citet{kim2018EmpathD} & XR Headset & 2 & no & no & no \\ \hline
        \textbf{Ours (\VIPSIM)} & \textbf{Desktop PC (Applications)} & \textbf{21} & \textbf{yes} & \textbf{yes} & \textbf{N=7 Iterations of the MonoWitness Protocol} \\ \hline 
    \end{tabular}
    \label{tab:comparison}
    \Description{This Table summarizes various publications related to visual impairment simulations categorized by the type of material used (e.g., special glasses, desktop PCs, AR/VR/XR headsets). It outlines the number of impairments simulated, the use of eye-tracking, involvement of visually impaired participants (VIPs), and whether user-centered design (UCD) methodologies were employed. The table highlights the diversity in technological approaches and levels of participant involvement, with the final row presenting the “Ours (VIP-Sim)” approach, which features the highest number of impairments (21), includes eye-tracking and VIP involvement, and is developed using 7 iterations of the MonoWitness Protocol.}
\end{table*}
\end{small}

\section{The False (?) Promise of Empathy}
The use of simulators also raises ethical questions. This section discusses the lack of inclusion and how simulators might reinforce such behavior.
\label{sec:promise}
\subsection{Use Cases of Vision Impairment Simulators}
%Zaman et al. presented an overview of visualizations in mixed reality for ophthalmic use cases. Categorizing into vision rehabilitation, vision assessment, and vision simulation. Vision simulation can further be separated into the use cases awareness, training (clinical or surgical), and accessibility (UIs or physical interaction)~\cite{zaman2024Advanced}.
Simulators offer potential benefits in the field of accessible design. Choo et al. evaluated Empath-D~\cite{kim2018EmpathD}, demonstrating that designers utilizing the simulator while redesigning Instagram gained a nuanced understanding of usability challenges faced by users with impairments, thus complementing existing guidelines-based approaches for general accessibility~\cite{choo2019Examining}. Tigwell et al., through designer interviews, identified educational value, resource efficiency, and improved design reliability as advantages of simulators~\cite{tigwell2021Nuanced}. Pratte et al. posit that simulators may enhance understanding, improve design solutions, and raise awareness, mainly when employed as training tools~\cite{pratte2021Evoking}. 

\subsection{Lack of Inclusion in Development}
While some works in \autoref{sec:rw_vissim} did include the opinion and guidance of VIPs during development (\citet{krosl2020CatARact} included cataract patients that received treatment, \citet{jones2020Seeing} state that a VIP was involved, \citet{giakoumis2014Enabling} state VIPs were involved), the majority (20/24) (see \autoref{tab:comparison} and \autoref{apdx:overview}) did not. In addition, the methodology (how VIPs could be included) and results remain unclear. This aligns with the findings by \citet{kasowski2023Systematica}. They conducted a systematic review of XR to understand and augment vision loss (n=277 publications) and concluded that it remains unclear if the simulations match the visual experience of VIPs. Only a limited number (n=2) of studies have firmly based their simulations on actual patient data~\cite{kasowski2023Systematica}.
It also remains unclear how to evaluate a vision impairment simulator with VIPs~\cite{schulz2019Frameworka}. The exclusion of VIPs might lead to false interpretations and underestimation of impairments. Therefore, we defined a methodological approach called the \textit{MonoWitness protocol} that includes VIPs to create \VIPSIM.

\subsection{A Fine Line Between Right and Wrong}
Despite potential benefits, the literature also highlights potential drawbacks. The replacement of people with disabilities by non-disabled individuals has long been a topic of discussion, as excluding the disabled community for the sake of convenience can be seen as disrespectful~\cite{bedrosian1995Limitations, mankoff2005Evaluatinga}. Simulators have been associated with the exclusion of authentic voices (people who are affected by impairments) and the evocation of discomfort~\cite{pratte2021Evoking}. This argument aligns with research describing simulators as potentially reinforcing negative attitudes and stereotypes toward individuals with impairments. By introducing the fear of becoming blind and, therefore, discomfort, these simulators can make accessible design seem like an unpleasant topic rather than a desirable goal~\cite{maher2024Stop, nario-redmond2017Crip, bennett2019Promise}. Users of disability simulations might misrepresent the true experiences of people with disabilities, leading to inaccurate perceptions among them. Designers using these simulations may focus on their perspectives rather than those of disabled individuals, and simulations frequently reinforce negative stereotypes rather than addressing the real social and infrastructural challenges~\cite{maher2024Stop, mattelmaki2014What, bennett2019Promise}. In addition, reinforcing ableism by professional expertism overriding authentic lived experience~\cite{leo2016Simulating} and oversimplifying complex conditions~\cite{geddes202330} might occur. Concerns regarding lack of realism and applicability have also been identified~\cite{dong2024Review}. We therefore assessed whether participants felt that \VIPSIM could replicate their symptoms.
It is noteworthy that the works mentioned above also focus on simulations of motor or cognitive disabilities. Furthermore, the interviewed groups vary, frequently comprising designers, teachers, students, and individuals with impairments. The interviews also often do not mention direct interaction with the simulator. To address potential concerns and not neglect the reasons for a negative toward simulators, we also present findings on the perception of \VIPSIM by VIPs when used by designers.

In conclusion, it is essential to emphasize that we do not advocate excluding VIPs from the development process of (software) products. Tools like Space$_x$Mag \cite{islam2023spacex} (in this case, a screen magnifier) provide assistive technology solutions for designs that have already been created but remain inaccessible. A simulation tool may prove beneficial, particularly in the early stages of development, where numerous decisions are made that are becoming increasingly costly to modify later. As Sears et al. suggest, using non-representative users is only appropriate for preliminary evaluations~\cite{sears2011Representing}. Also, as VIP feedback is a valuable resource, it might be wise to use a simulator to identify and mitigate minor issues in advance.
We propose that when used with caution, simulators can offer advantages to both the user of simulators and those affected by impairments, particularly regarding educational value for designers and awareness during the prototyping phase~\cite{mankoff2005Evaluatinga}.

\section{Methodology}
%\subsection{Participatory Design Including People With Vision Impairment}
We now introduce our UCD process to include VIPs in developing and evaluating \VIPSIM.

\subsection{On the Challenge of Including People with Vision Impairments in the Design Process}
Many works do not report how their simulation's design and implementation choices were made (see \autoref{sec:rw_vissim}). The textual description of other related works or descriptions of medical experts justified some design choices~\cite{kasowski2023Systematica}.

%Regarding evaluating the simulators, \citet{jones2020Seeing} report testing whether their simulator could functionally approximate the actual experience of the glaucoma patient by testing whether the simulation also led to longer search times when tasked to find a smartphone.

However, an evaluation of simulations with VIPs is missing; such assessments as done by \citet{jones2020Seeing} or \citet{krosl2023Exploring} only capture a small part of the experience of VIPs, as they often have more than one symptom. \citet{krosl2020CatARact} focused on patients who received treatment and are already cured of cataracts. 
A particular challenge is that VIPs will perceive a layering of effects when using \VIPSIM: (1) the effects of their impairment and (2) the simulation of the impairment (see \autoref{fig:meth}). 

Therefore, we devised a novel methodology to define, implement, and validate the vision impairment simulation called \textit{MonoWitness Protocol}. 
Our method aims to reduce such layering effects and be personalized to the individual impairment. 
The method included an initial development and iterative refinement of \VIPSIM using the novel \textit{MonoWitness Protocol}.

%After developing our initial prototype, we used this procedure to ask VIPs their opinions regarding \VIPSIM. The following section describes our process to receive and incorporate feedback on \VIPSIM. 

\subsection{Initial Development}
Visual impairments and their symptoms have a wide range of effects. Therefore, we built on previous work and first implemented a set of symptoms for our initial version of \VIPSIM, before conducting evaluations with VIPs. These symptoms were chosen due to their prevalence, as reported by the WHO~\cite{stats2023who}, through a literature review of \citet{zaman2024Advanced} and OpenVisSim~\cite{jones2020Seeing}.

%Then, \VIPSIM underwent a development process where every participant evaluated the current version and incorporated changes depicted in \autoref{fig:process}.

\begin{figure*}[ht]
    \centering
    \includegraphics[width=\textwidth]{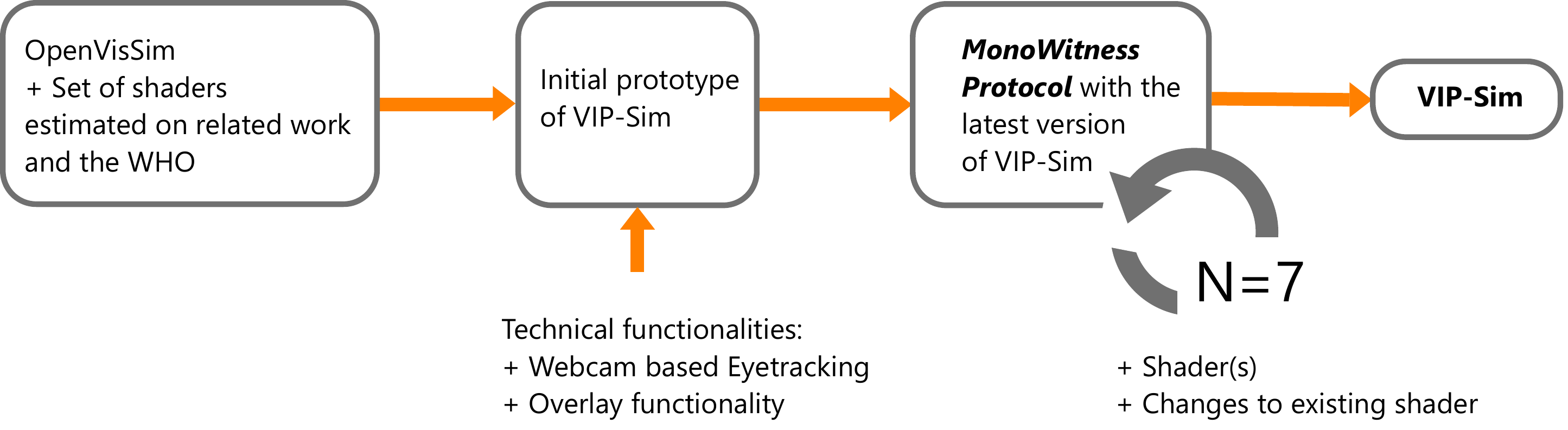}
    \caption{Overview of the development process of \VIPSIM. Starting from an unevaluated estimate of the shader for visual impairments, we underwent the \textit{MonoWitness Protocol} seven times to evaluate, refine, or extend it.}
    \Description{A flowchart illustrating the development process of VIP-Sim. It starts with OpenVisSim, which includes a set of shaders based on related work and WHO standards. From this, an initial prototype of VIP-Sim is developed. Next, technical functionalities such as webcam-based eye tracking and overlay functionality are added. The MonoWitness Protocol is then applied using the latest version of VIP-Sim. This cycle repeats with N=7 iterations, involving new shaders and modifications to existing shaders, leading to the final version of VIP-Sim.}
    \label{fig:process}
\end{figure*}

After this initial implementation, we included VIPs in the evaluation and iterative refinement of \VIPSIM via the \textit{MonoWitness Protocol} (see: \autoref{fig:meth}). 
Every participant evaluated the current version \VIPSIM. We then incorporated changes before the next participant evaluated the then up-to-date \VIPSIM (see \autoref{fig:process}).

\subsection{Novel Method -- MonoWitness Protocol}
\label{sec:MonoWith}

\begin{figure*}[ht]
    \centering
    \includegraphics[width=\textwidth]{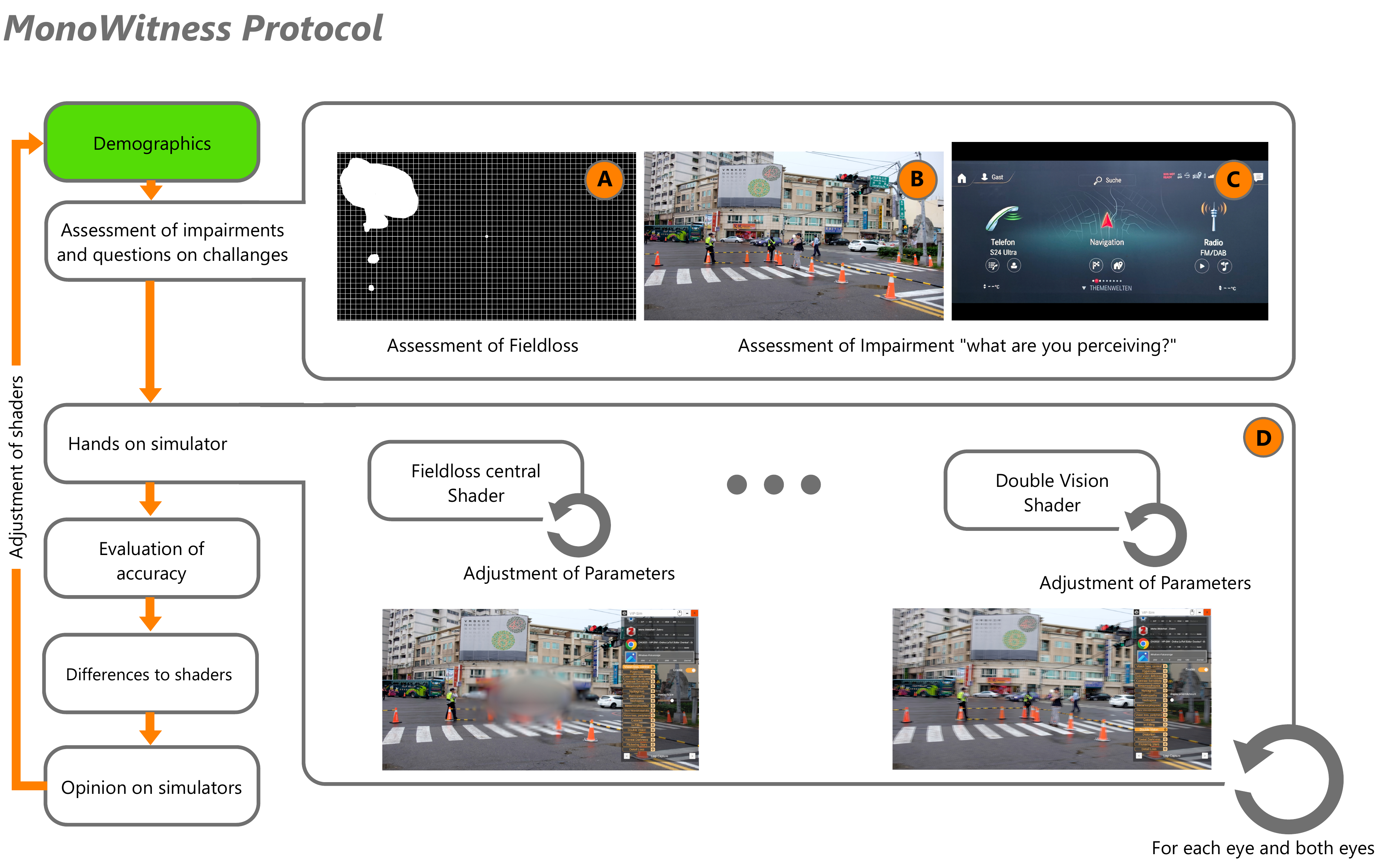}
    \caption{Our approach for assessing and participatory development of the visual impairment shaders. The left-hand side depicts our iteration cycle over all participants. We first assess the participant's vision and the challenges (A-C). Next, a hands-on session (D) was held to adjust the shader parameters. (A) depicts the Amsler grid \cite{amsler_grid_wikipedia} test implemented in \VIPSIM. (B) Participants were asked to describe the first scene incorporating a contrast sensitivity \cite{pelliDesigning1987} and color deficiency \citet{ishihara1951tests} test. (C) Similar to (B), participants had to describe their perceptions and how their impairments affected them. Then, the study supervisor and the participant reviewed the current version of \VIPSIM according to the \textit{MonoWitness protocol} (D). The shaders were set to their impairment (by the VIPs or if requested by the supervisor). If a shader was insufficient or missing to represent their perception, the differences or new shaders were implemented before the next participant.}
    \Description{This figure shows a diagram of a process. On the left-hand side, a loop connects several boxes with text inside. The first box, located at the top left, is labeled "Demographics". Below it is another box labeled "Assessment of Impairments and Questions on Challenges", which folds upwards and to the right. Inside this box are three images: the first is a depiction of an Amsler grid with drawings on it labeled "A". The second is a depiction of a city traffic scene, labeled "B," The third is an image of a car's user interface, labeled "C". Below the Amsler grid is the caption "Assessment of Field Loss." In contrast, below the city scene and car UI, the caption reads "Assessment of Impairment: What are you perceiving?" The next box is labeled "Hands-On Simulator". In the bottom right corner of this box is an icon representing "Repeat," labeled "For each and both eyes". This box also folds upwards to the right. Inside this section, labeled "D," are two smaller boxes. The first box on the left is labeled "Field Loss Central Shader". The same repeat icon appears here, accompanied by the "Adjustment of Parameters" label. Below this box is a depiction of the city scene and the VIP-Sim UI, where the center of the image becomes very blurry. The second smaller box within the "Hands-On Simulator" section is similar to the first but labeled "Double Vision Shader". In this box, the image of the city scene below is doubled with an offset. Three large dots between these two smaller boxes indicate that other shaders are being tested in between. After the "Hands-On Simulator" box, there is a box labeled "Evaluation of Accuracy," followed by another box labeled "Differences to Shaders" and "Opinion on Simulators". Finally, an arrow loops back to the "Demographics" box, labeled "Adjustment of Shaders."}
    \label{fig:meth}
\end{figure*}

\citet{taylor2018Seeing} let people with AMD verbally describe their symptoms after an eye examination and assess a common picture depicting AMD. However, this was tailored for AMD and lacked specificity due to the verbal description. \citet{krosl2020CatARact} demonstrated that this approach works for cataracts where one eye is treated. We extended these methods by incorporating vision tests, requesting verbal descriptions, and finally presenting our initial shaders as a starting point to identify differences in the participants' perceptions of \VIPSIM's representation of the impairment.

Our approach, the \textit{MonoWitness protocol}, primarily relies on semi-structured interviews with a hands-on section on the simulation by \VIPSIM. This procedure (interviews and hands-on sections) is widely used in accessibility research~\cite{brewer2018Understanding, colley2020Inclusive, 10.1145/3546717, 10.1145/3659618, 10.1145/3706598.3713454}.

The \textit{MonoWitness protocol} is designed to bridge the gap between subjective visual experience and its digital simulation. The \textit{MonoWitness protocol} combines:
\begin{itemize}
    \item \textbf{Calibration of the Simulation}: Using a custom assessment tool (detailed in \autoref{sec:assessment}) to match \VIPSIM with each participant’s unique visual profile.
    \item \textbf{Hands-On Simulation and Comparison}: Participants view personalized simulations and alternate between their left and right eyes. This step leverages the differences in the vision of our participants' eyes.
    \item \textbf{Semi-Structured Interviews}: Participants describe their visual experience both verbally and through interactive comparisons to identify differences.
    \item \textit{(Changes to \VIPSIM)}: Based on the identified differences, we adjusted \VIPSIM iteratively.
\end{itemize}

First, we created an assessment tool (see \autoref{fig:meth} A-C) to allow the study supervisor to calibrate \VIPSIM to match the perceived perception of the participant quickly. Based on this assessment, we showed the participants an initial personalized view using \VIPSIM. Then, the hands-on simulator section of the MonoWitness protocol was used to calibrate the existing shaders. The calibration was done by comparing them (with the VIPs) to the VIPs' described perception and determining if and where changes in \VIPSIM needed to be made to represent their perception. This was particularly challenging. Describing what is seen can be straightforward, but addressing what remains unnoticed presents a unique difficulty. Although this may seem counterintuitive at first, our participants were still able to effectively utilize their remaining vision, as follows:

Our innovation lies in primarily recruiting participants who once had a vision impairment and were cured or who have differing eyesight between eyes (e.g., one eye has a vision impairment while the other does not). This is represented in our method's name: MonoWitness, as participants had witnessed the impairment or had it in one (mono) eye.
The participant first looked with the left eye while covering the right, and then with the right eye while covering the left. We did this because of both \textbf{symptoms and their effects on the sight variety between eyes}. 
For example, Participant 3 (P3) had peripheral vision loss in the left eye and foveal vision loss in the right eye. We showed them our visual simulation for foveal vision loss while using the right eye. This allowed them to compare their actual visual impairment to the simulated effects and evaluate the shader. Since their right eye had only foveal vision loss, they could still see through their peripheral vision and observe what the peripheral vision loss shader looked like, and vice versa. This procedure was so effective that (P3) could adjust and scale the size of the simulated vision loss on their own. (P1) used their healthier eye (with only mild clouding) to describe what happened to their sight with their impaired eye by swapping between the left eye and the severely impaired right eye.
Some visual disturbances like teichopsia can occur sporadically~\cite{rasmussen1992migraine}, and some impairments, like cataracts, can be treated by, e.g., a lens replacement, which was the case for (P1) and (P4).
These participants were especially invaluable to us as they could both describe how their vision was affected when they had the impairments and how they \textbf{currently} perceive the world. 

Not all participants had differences in sight between their eyes (e.g., (P5) and (P6)). This emphasizes that the \textbf{adjustments and protocol must be personalized}. In these cases, other compensations were applied. Participants scanned the image and took extra time to perceive what was shown. Additionally, screen magnification, which displayed parts of the simulation in a larger format, enabled them to observe and investigate what was shown on the screen.  While these alternative viewing methods likely resulted in reduced efficiency compared to direct perception, our results indicated that participants were still able to comprehend and work with the displayed information.

Finally, we allowed participants to use their \textbf{binocular vision} to identify visual differences. While the monocular vision helped us get an unconfounded impairment description (e.g., if just one eye is impaired), for our individuals, binocular vision was superior as their eyes compensated for each other’s deficits, providing a more comprehensive perception of the visual effects.

The following questions (translated from German) were asked before the \textit{MonoWitness Protocol}. We asked if participants had a medical diagnosis of their impairment, were receiving treatment, and whether participants were currently affected by their impairment. 

\begin{enumerate}
    \item[Q0] Do you have a medical diagnosis for your impairment?

    \item[Q1] Are you currently affected by your impairment, if not, when was the last time you were affected?

    \item[Q2] Have you ever been treated for your impairment? Are you currently being treated for your condition? Will you seek treatment for your condition in the future?

    \item[Q3] How does your impairment affect your use of a technical device, such as a computer, TV, or smartphone?

    \item[Q4] What strategies do you use to mitigate the effects of your impairment (coping strategies)?

    \item[Q5] What aids do you use to mitigate the effects of your impairment?

\end{enumerate}

These descriptions helped to set the initial state of \VIPSIM more precisely.
During the \textit{MonoWitness protocol}, these questions were asked:

\begin{enumerate}
    \item[Q6] Here is a picture of a landscape. If possible, please try to fixate the marked point in the center. Can you describe what you see?

    \item[Q7] If possible, please draw which part of the image is affected and how you perceive the affected areas?

    \item[Q8] You will now see visual effects on the screen. These visual effects have different effect sizes that we can adjust together. Please describe how applicable the effects are in comparison with your changed vision due to the impairment.
\end{enumerate}

These questions were asked at the end of the session:

\begin{enumerate}
    \item[Q9] Imagine a designer uses a tool to simulate the symptoms of your vision impairment, how are your feelings about that?
    
    \item[Q10] Do you have any concerns regarding designers using simulators while designing?
\end{enumerate}

\subsubsection{Assessment Session}\label{sec:assessment}
The participants sat in front of a 27-inch 2440x1440 monitor (Asus VA27AQSB). During the tests, a monitor-to-eye distance of 60cm was ensured. The seat's height was also controlled so that the height of the eyeline was equal to the center of the screen. The lighting conditions were also controlled by closing the blinds. After the initial assessment, participants could move closer to the screen, especially when their visual acuity was insufficient.

Our assessment tool for \VIPSIM was created using Unity 2022.3.27f1. The assessment tool consists of three parts. First, it displays an Amsler grid (see \autoref{fig:meth} A). The Amsler grid is a test to assess vision field loss in the central vision. In the traditional test, a 10cmx10cm grid is printed, placed, and held in front of the face so that each grid cell covers 1° of the central vision~\cite{amsler_grid_wikipedia, wang2024How}. We inverted the color of the grid, as it is more pleasant to look at for people sensitive to light. Using the screen's dimensions and pixel density, the tool adjusts the cell size based on the distance to the screen. Participants were asked to fixate on a point in the center of the screen and keep their heads at a constant distance from the screen. %place, e.g., not moving closer to the screen. 
The tool allows us to draw and mark the locations of symptoms of visual impairments. We located these while asking the participants how they perceived the lines. (P1, P2, P4), and (P6) (see \autoref{tab:participants}) drew their field loss personally on the grid.

The tool also allowed displaying two high-resolution (2560px $\times$ 1440px) images (see \autoref{fig:meth} B) that depict a crowded city street scene (Unique Colors: 493010, Complexity: 1642KB \cite{marin2013examining}) and the interface of a 2024 Mercedes-Benz E-Class All-Terrain (Unique Colors: 150176, Complexity: 783KB \cite{marin2013examining}) see \autoref{fig:meth}. We used this scene because of the contrast in complexity and use case. We edited the city scene by replacing a billboard in the center of the image with two vision tests. On the left is a rebuild of the Pelli-Robson chart~\cite{pelliDesigning1987}. On the right are two Ishihara test plates~\cite{ishihara1951tests} (Plate No. 13 and Plate No. 23) to receive an estimate of contrast sensitivity and possible (undiagnosed) color blindness. During the test, specific impairments may have interfered with each other. For example, the contrast sensitivity test was not feasible due to insufficient visual acuity. To address this, we allowed the use of screen magnification. This ensured that the letters retained the same contrast level for the contrast sensitivity test but appeared larger.

With our participants, we first assessed areas affected by the impairment and marked them on the Amsler grid. In the next step, both images were shown, and participants were asked to focus on the point at the center of the picture. Then, the participants were prompted to describe what they saw in the image and how the affected areas were perceived. This process was repeated for each eye individually and then for both eyes to get more insights since participants' acuity varied between their eyes.

\begin{figure*}[ht]
    \centering
    \includegraphics[width=\textwidth]{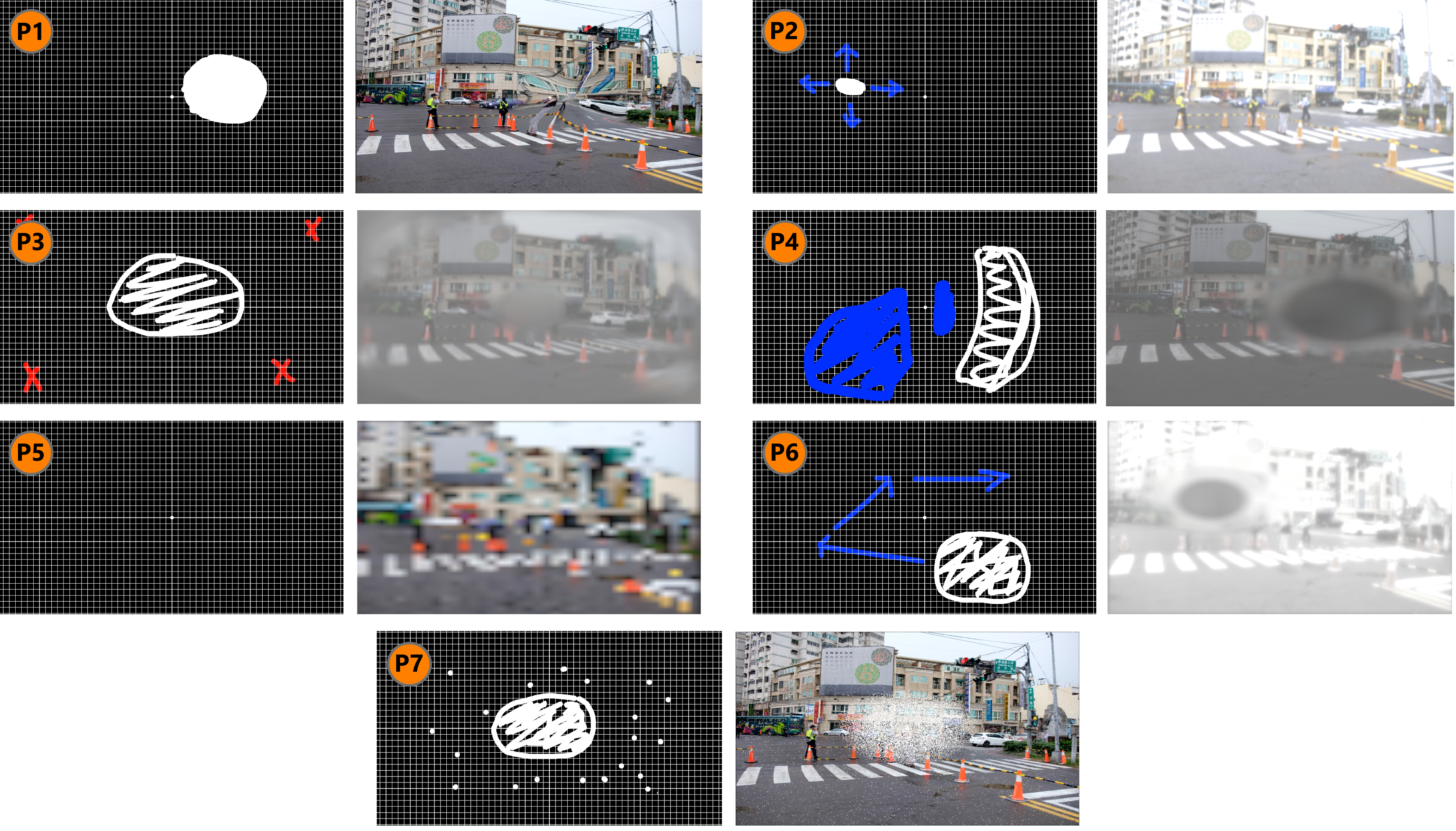}
    \caption{Illustration of the impairments identified with our participants. The marked areas on the left indicate regions with visual field loss. The impairments simulated by \VIPSIM are displayed on the right side. It is important to note that the visual field deficits dynamically move in sync with the gaze, as tracked by the eye-tracking system.}
    \label{fig:assess}
    \Description{The illustration presents a side-by-side comparison of the field loss assessment, drawn on a grid, with the simulated impairments experienced by each participant. All participants, except P5, have circular drawings on the grid. The image on the right shows a city scene. For P1, a suction effect pulls pixels toward the center, creating a vortex-like appearance. For P2, the image appears brighter and blurry, with a grey spot in the center. For P3, the image's outer and inner parts are blurred, with a grey overlay, further increasing the blurriness. For P4, the image is blurry, with a dark, blurry spot in the center. P5's image is heavily pixelated. P6's image is extremely bright, almost white, with a dark spot visible. Finally, P7's image shows white spots scattered throughout, forming a central cluster.}
\end{figure*}

Then, \VIPSIM was used on the images used for assessment (see \autoref{fig:assess}). The eye-tracking feature (realized via UnitEye~\cite{wagner2024UnitEye}) of \VIPSIM was enabled. Eye-tracking was not always feasible due to an eye condition like nystagmus (this occurred for (P5)), where participants showed jumping eye movement~\cite{rosengren2020Robust}. We, therefore, prepared a fallback feature to attach the point of gaze to the mouse pointer. In this case, participants then showed us how their gaze moved, and the visual effects interacted by dragging the mouse on their own. However, for the end-users (designers), eyetracking is crucial for understanding this behavior. For each available symptom and shader, participants were asked if the simulated vision was similar to their perceived vision. If shaders did not reflect the impairment (e.g., for a participant without Color Vision Deficiency (CVD), the CVD shader), they were not shown as they did not yield meaningful data. Differences in the simulated depiction were then used with the image description to estimate where the simulations needed to be adjusted. If possible, the study supervisor adjusted them during the study via \VIPSIM's settings (participants could set the sliders themselves). Otherwise, we asked for a description in addition to the descriptions given during the assessment and implemented it before the next participant was interviewed. Therefore, we could produce subjective approximations of symptoms, which, as our results show, are generalizable among our sample. A detailed description of our symptom visualizations is described in \autoref{sec:shaders}.

\section{Study and Iterative Refinement of \VIPSIM}
%After having described previous vision impairment simulations and our initial prototype of \VIPSIM, which relied heavily on prior work, we now introduce our methodology to improve the vision impairment simulation of \VIPSIM by including VIPs.

%This chapter introduces our study, the UCD process, and the final version of \VIPSIM.
Our study was motivated by the research question (RQ):

\begin{quote}
\researchq{RQ1}{How well does \VIPSIM simulate vision impairments?}
\end{quote}

\label{sec:meth_study}

\subsection{Procedure}
A consent form was completed after the participants were introduced to the survey topic and its procedure. The researcher offered to read the participant's consent form. Participants were compensated with 13€. The study and the process followed all institutional ethical guidelines. Furthermore, participants could withdraw from the study at any time and take breaks to relieve their eyes as needed. Additionally, participants were permitted to wear eye protection glasses at any time. An interview session took between 35 and 97 minutes, M=50.57, SD=19.33. We recorded the entire interview, and all data is publicly available (on GitHub) in anonymized form. In the first part of the interview, demographic information was collected, especially about their impairments. Then, the assessment occurred (see Section~\ref{sec:assessment}). Finally, participants evaluated the vision impairment simulation and were asked the questions of the \textit{MonoWitness Protocol} (see \autoref{sec:MonoWith}).

\subsection{Participants}
We interviewed N=7 VIPs (see \autoref{tab:participants}). Participants were recruited using notices or callouts in groups or associations with a mean age of M=54.86 (SD=20.52). 4 identified themselves as male and 3 as female.

\begin{table*}[ht]
    \footnotesize
    \begin{tabular}{c|c|c|>{\RaggedRight\arraybackslash}p{2cm}|>{\RaggedRight\arraybackslash}p{3cm}|>{\RaggedRight\arraybackslash}p{3cm}|>{\RaggedRight\arraybackslash}p{2cm}}
    %\hline
    \textbf{ID} & \textbf{Age} & \textbf{Gender} & \textbf{Visual Acuity (in percent)} & \textbf{Diagnosis} & \textbf{Description} & \textbf{Onset} \\
    \hline
    P1 & 77 & F & unknown & AMD & wet AMD on left eye dry AMD on right eye & 2 years \\
    \hline
    P2 & 60 & M & left eye 0.1 right eye 0.3 & optic atrophy, nystagmus, axis deviation, nearsighted, sensitive to light, retinal opacities, vitreous opacity & blurry vision, floating intraocular particle & birth (degenerative)\\
    \hline
    P3 & 29 & M & 0.15 for both eyes & retinitis pigmentosa (RP) & foveal and peripheral visual restriction & birth (degenerative) \\
    \hline
    P4 & 68 & F & 0.1 left eye, 0.1-0.15 right eye & autosomal dominant drusen & foveal vision loss & 22 years\\
    \hline
    P5 & 62 & M & 0.005 left eye, 0.05 right eye & cataract, nystagmus and others (unknown) & loss of detail perception, only color areas are visible if not in immediate proximity. & birth\\
    \hline
    P6 & 65 & M & 0.03 both eyes & cone-rod dystrophy, optic atrophy & blurred vision, foveal visual restriction & 22 years\\
    \hline
    P7 & 23 & F & 0.05 left eye 0.08-0.09 right eye & optic atrophy & white patches and dots & about 10 years\\

    \end{tabular}        
    \caption{Demographic information about VIPs that took part in the survey and development process of \VIPSIM}
    \Description{The table presents medical data for seven participants with visual impairments. The columns include ID, age, gender, visual acuity (in percent), diagnosis, description of the visual condition, and onset. Ages range from 23 to 77 years. Diagnoses include AMD (age-related macular degeneration), optic atrophy, retinitis pigmentosa, cataracts, cone-rod dystrophy, and autosomal dominant drusen. Visual acuity varies significantly, with some individuals having as little as 0.03 percent vision, and one case with unknown acuity. Descriptions of the visual impairments include foveal and peripheral vision loss, blurred vision, sensitivity to light, floating particles, loss of detail and color perception, and presence of white patches. The onset of the condition ranges from birth (degenerative) to adult onset, such as 2 or 22 years ago.}
    \label{tab:participants}
\end{table*}

% Previous work has shown that the simulation of impairments can increase empathy~\cite{krosl2023Exploring}

\subsection{Introducing \VIPSIM: Visual Impairment Simulator}
Our UCD process led to the development of 21 symptom shaders. Each shader was tailored to simulate the unique characteristics of these visual impairments, informed by user feedback.

Using \VIPSIM, the user can select the currently running software for which the impairment shall be simulated. We made the application transparent and click-through as if no application were running above. The user can, therefore, interact with the underlying application of their choice at any time. If the software is selected and the simulation is enabled, the transparent Unity overlay is applied, which depicts what the software shows with the applied symptoms. Therefore, the user can design while having a simulation and can directly see the changes in the design through a simulated impairment.
We hypothesize that this easy one-click toggle leads to more frequent use than other tools, where the validation breaks the current workflow. After selecting an impairment, the user can modify impairment-specific parameters such as severity. For impairments like AMD, (peripheral) field loss, or floaters (dots on the screen to simulate field loss), we employed eye-tracking using UnitEye~\cite{wagner2024UnitEye}. %The source code for \VIPSIM is available at \url{https://github.com/Max-Raed/VIP-Sim}.
\autoref{fig:ui} depicts our UI. \VIPSIM was implemented in Unity 6000.0.28f1. %Modifications were made for \VIPSIM to work in an overlay setting with only one camera for eye-tracking, and because of deviations with the participants' perceptions. 
% We conducted a literature review and found that these were used:
% Implementation details regarding Shaders and Overlay mechanics

% If there are limitations, what is important
% HERE: Thinking process on how the selected impairment was chosen.

\subsection{Shaders and Visualizations}
\label{sec:shaders}
\begin{figure*}[htbp]
\centering
\captionsetup[subfigure]{justification=centering}

% Erste Reihe (Bilder 1–4)
\begin{subfigure}[t]{0.24\textwidth}
    \includegraphics[width=\linewidth]{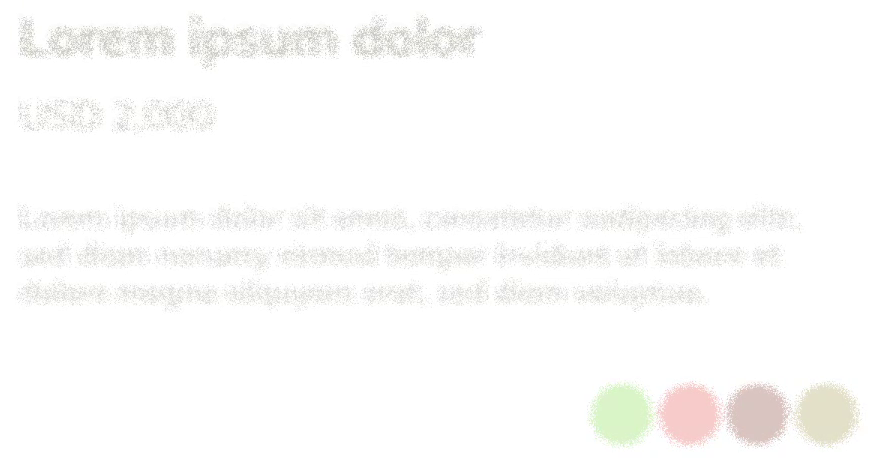}
    \caption{Cataracts}
    \label{fig:bild1}
\end{subfigure}
\hfill
\begin{subfigure}[t]{0.24\textwidth}
    \includegraphics[width=\linewidth]{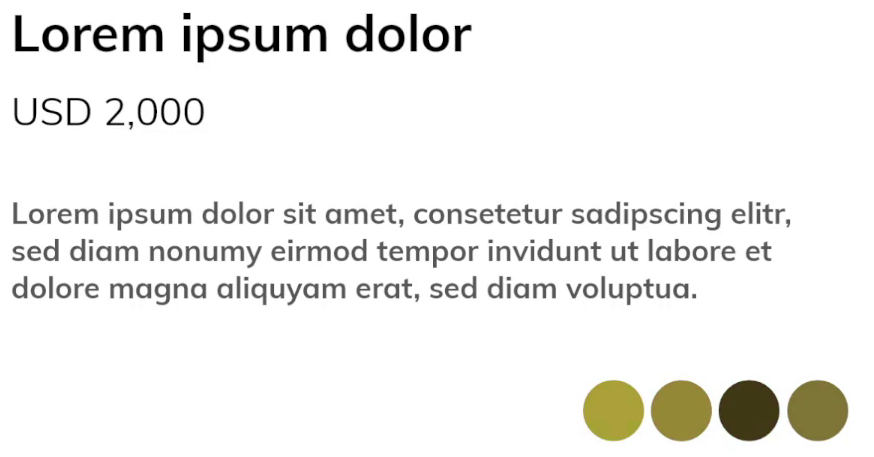}
    \caption{Color vision deficiency - Deuteranomaly}
    \label{fig:bild2}
\end{subfigure}
\hfill
\begin{subfigure}[t]{0.24\textwidth}
    \includegraphics[width=\linewidth]{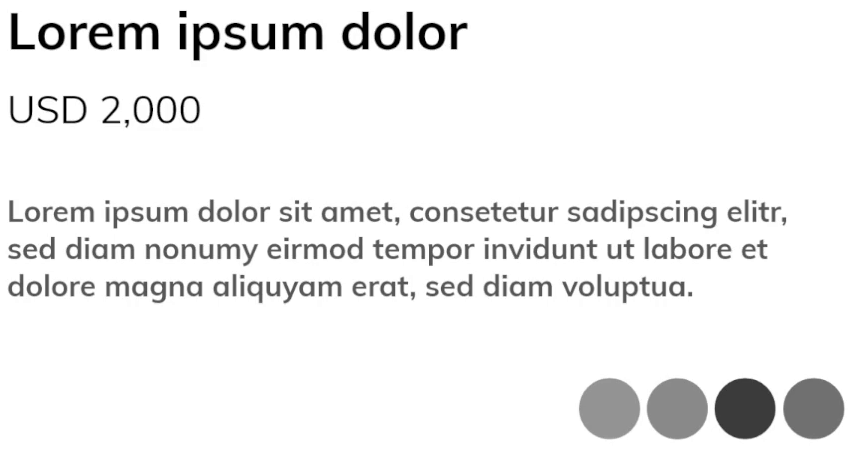}
    \caption{Color vision deficiency - Monochromia}
    \label{fig:bild3}
\end{subfigure}
\hfill
\begin{subfigure}[t]{0.24\textwidth}
    \includegraphics[width=\linewidth]{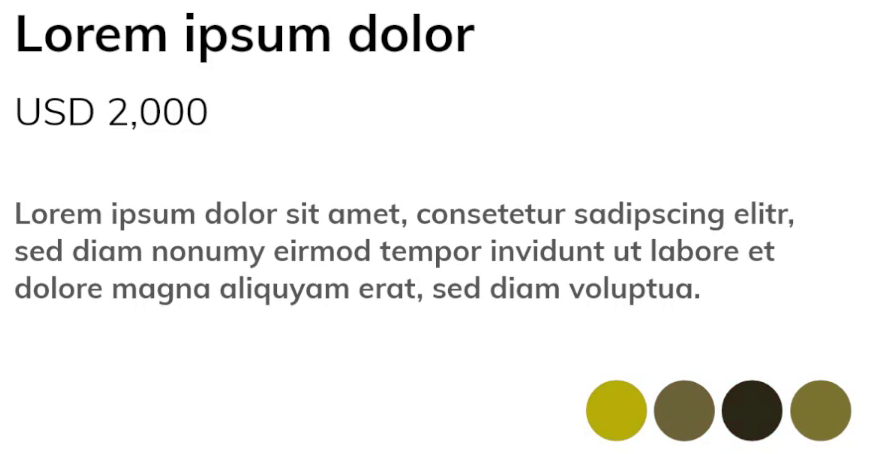}
    \caption{Color vision deficiency - Protanomaly}
    \label{fig:bild4}
\end{subfigure}

\vspace{0.5cm} % Abstand zwischen den Reihen

% Zweite Reihe (Bilder 5–8)
\begin{subfigure}[t]{0.24\textwidth}
    \includegraphics[width=\linewidth]{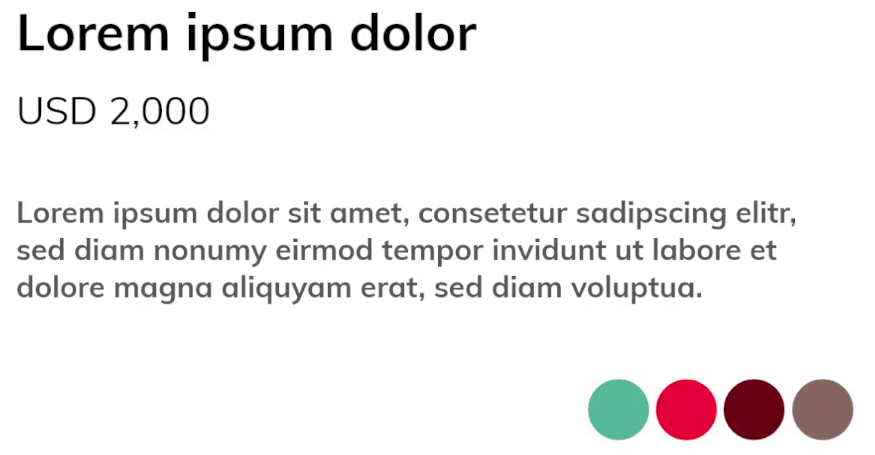}
    \caption{Color vision deficiency - Trianomaly}
    \label{fig:bild5}
\end{subfigure}
\hfill
\begin{subfigure}[t]{0.24\textwidth}
    \includegraphics[width=\linewidth]{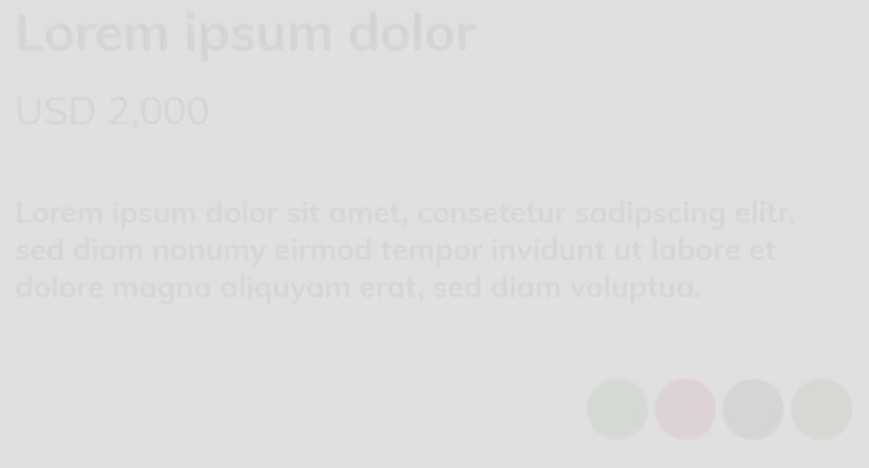}
    \caption{Contrast sensivity}
    \label{fig:bild6}
\end{subfigure}
\hfill
\begin{subfigure}[t]{0.24\textwidth}
    \includegraphics[width=\linewidth]{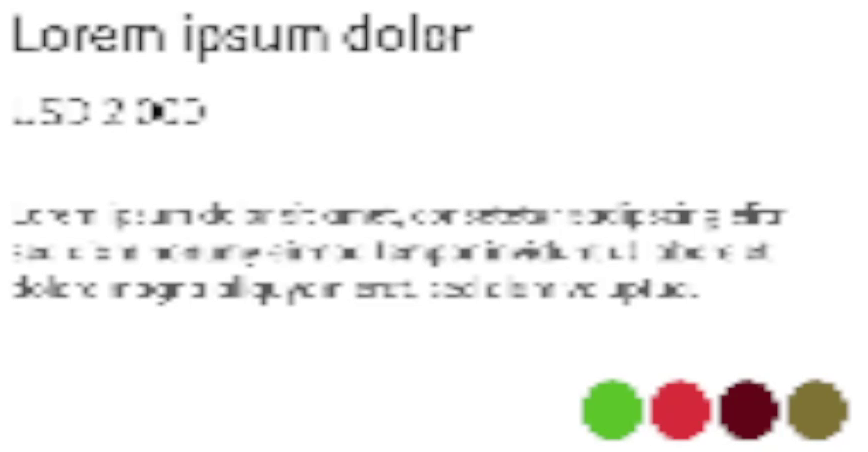}
    \caption{Detail Loss}
    \label{fig:bild7}
\end{subfigure}
\hfill
\begin{subfigure}[t]{0.24\textwidth}
    \includegraphics[width=\linewidth]{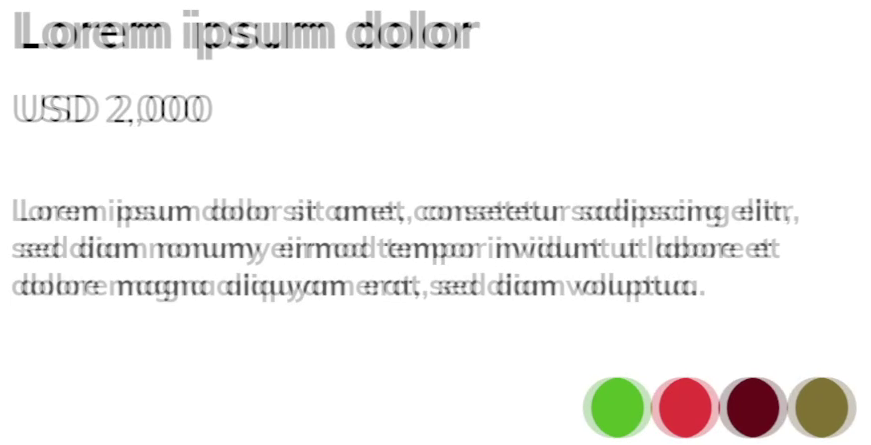}
    \caption{Double Vision}
    \label{fig:bild8}
\end{subfigure}

\vspace{0.5cm}

% Dritte Reihe (Bilder 9–12)
\begin{subfigure}[t]{0.24\textwidth}
    \includegraphics[width=\linewidth]{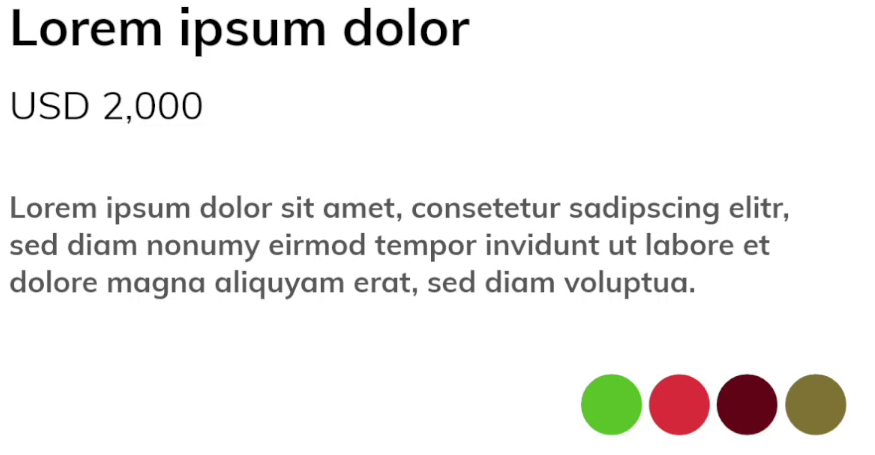}
    \caption{Flickering Stars}
    \label{fig:bild9}
\end{subfigure}
\hfill
\begin{subfigure}[t]{0.24\textwidth}
    \includegraphics[width=\linewidth]{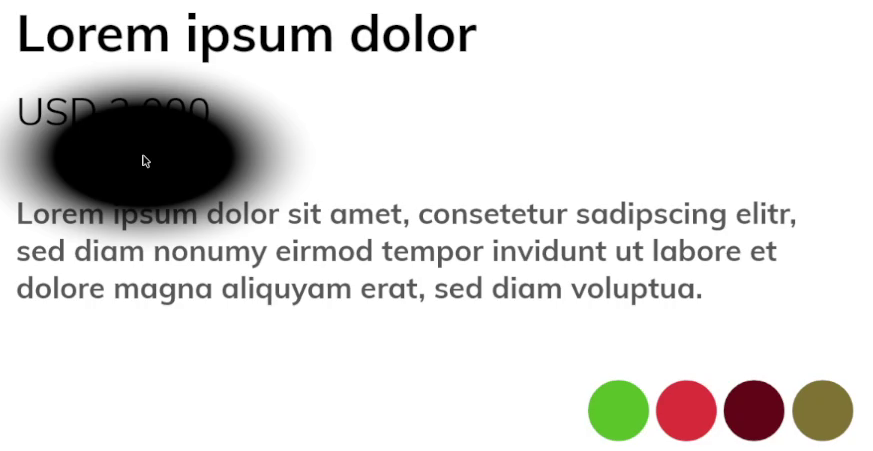}
    \caption{Foveal Darkness}
    \label{fig:bild10}
\end{subfigure}
\hfill
\begin{subfigure}[t]{0.24\textwidth}
    \includegraphics[width=\linewidth]{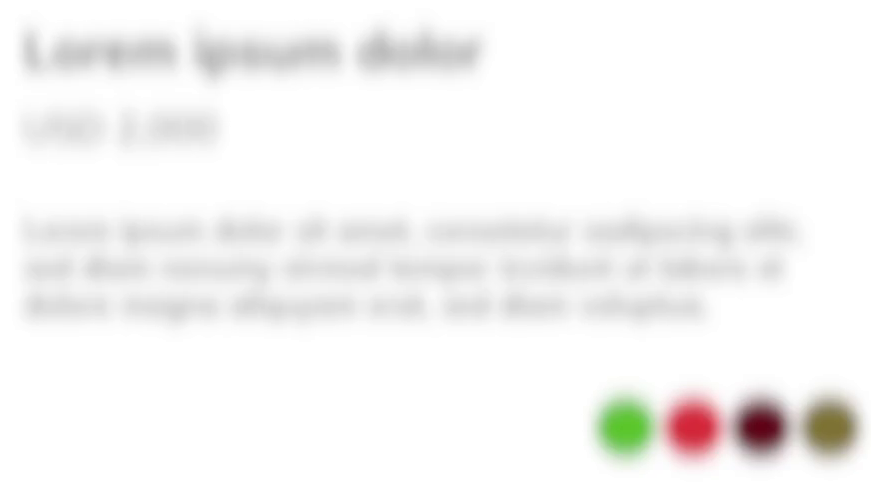}
    \caption{Hyperopia}
    \label{fig:bild11}
\end{subfigure}
\hfill
\begin{subfigure}[t]{0.24\textwidth}
    \includegraphics[width=\linewidth]{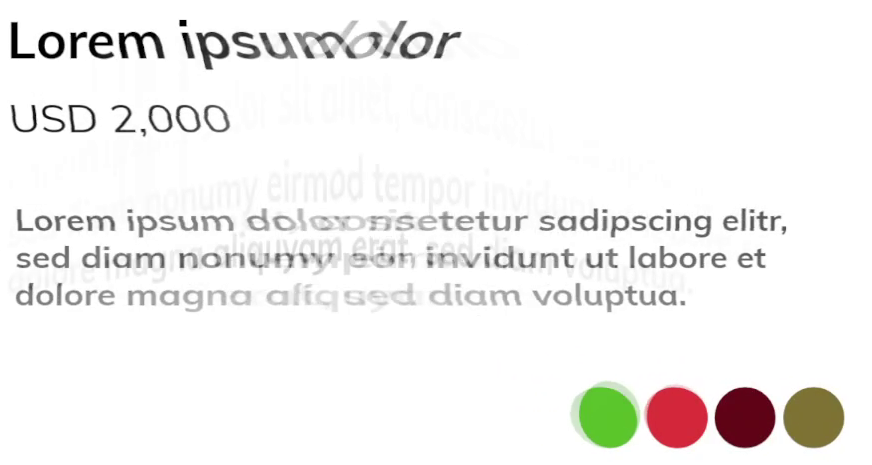}
    \caption{In-Filling}
    \label{fig:bild12}
\end{subfigure}

\vspace{0.5cm}

% Vierte Reihe (Bilder 13–16)
\begin{subfigure}[t]{0.24\textwidth}
    \includegraphics[width=\linewidth]{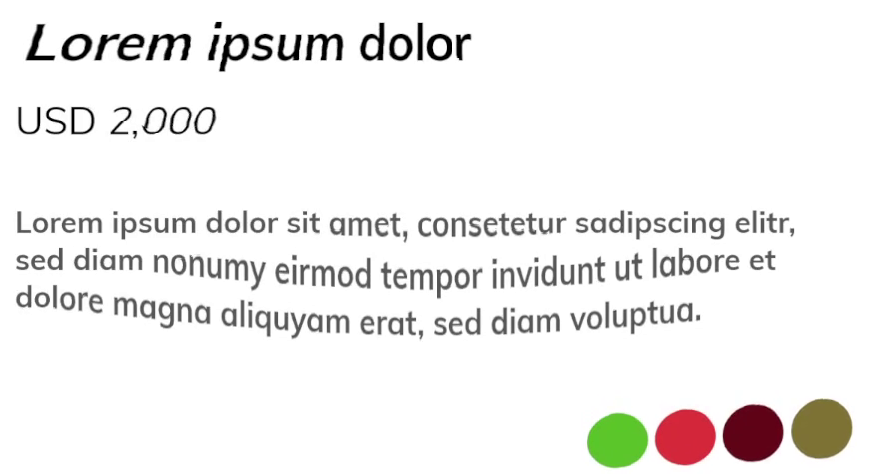}
    \caption{Metamorphosia}
    \label{fig:bild13}
\end{subfigure}
\hfill
\begin{subfigure}[t]{0.24\textwidth}
    \includegraphics[width=\linewidth]{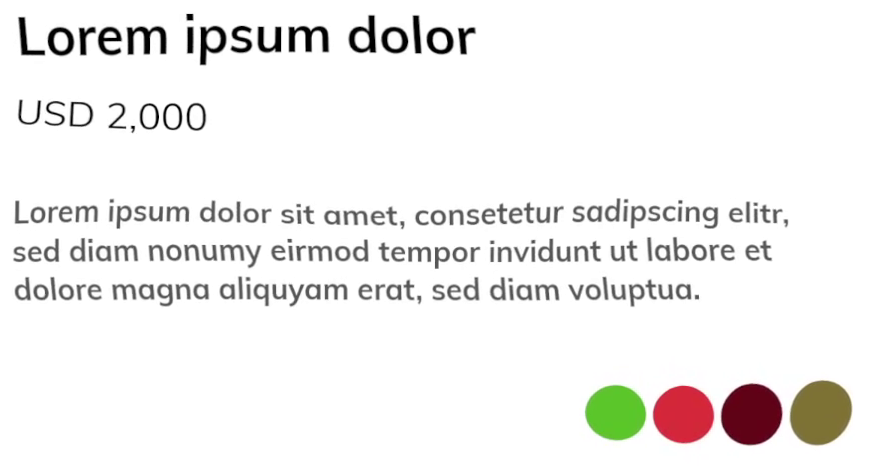}
    \caption{Metamorphosia - overlay}
    \label{fig:bild14}
\end{subfigure}
\hfill
\begin{subfigure}[t]{0.24\textwidth}
    \includegraphics[width=\linewidth]{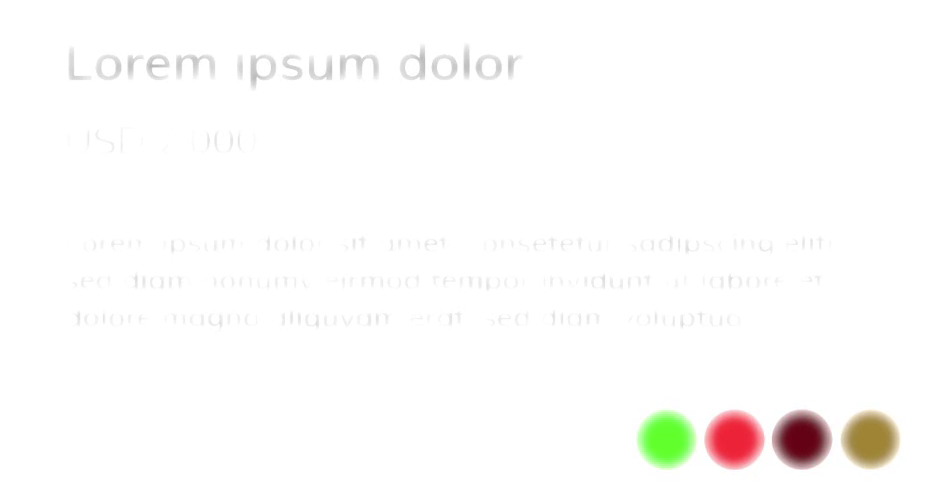}
    \caption{Glare Vision/Photophobia}
    \label{fig:bild15}
\end{subfigure}
\hfill
\begin{subfigure}[t]{0.24\textwidth}
    \includegraphics[width=\linewidth]{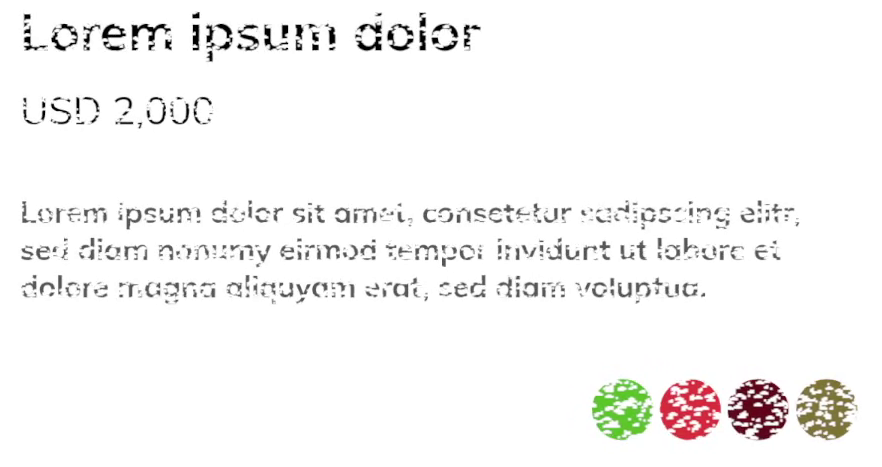}
    \caption{Retinopathy}
    \label{fig:bild16}
\end{subfigure}

\vspace{0.5cm}

% Fünfte Reihe (Bilder 17–20)
\begin{subfigure}[t]{0.24\textwidth}
    \includegraphics[width=\linewidth]{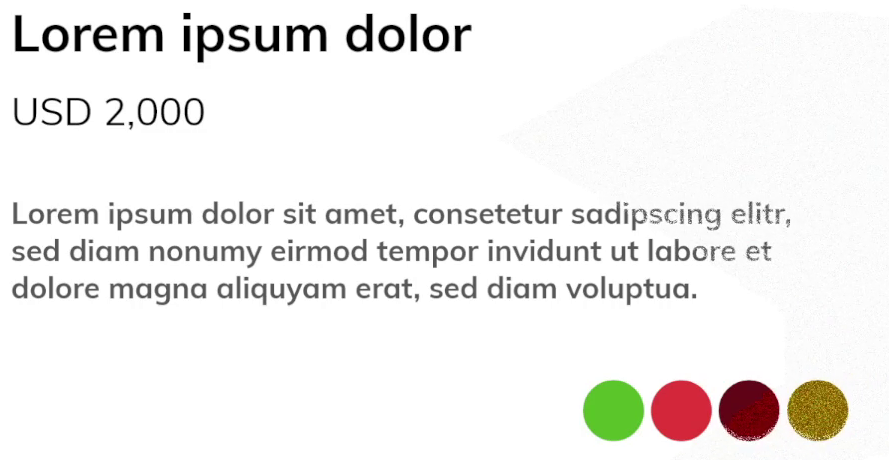}
    \caption{Teichopsia}
    \label{fig:bild17}
\end{subfigure}
\hfill
\begin{subfigure}[t]{0.24\textwidth}
    \includegraphics[width=\linewidth]{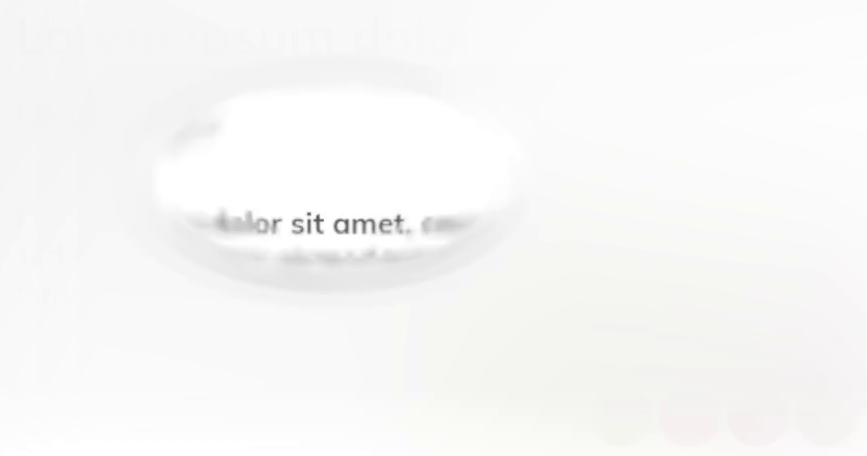}
    \caption{Vision loss, peripheral}
    \label{fig:bild18}
\end{subfigure}
\hfill
\begin{subfigure}[t]{0.24\textwidth}
    \includegraphics[width=\linewidth]{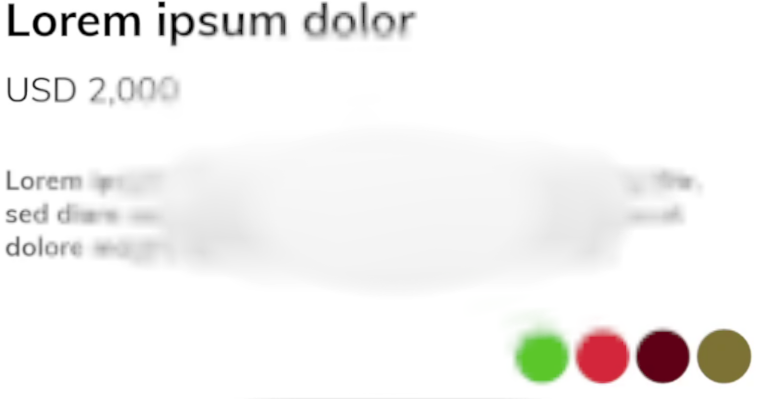}
    \caption{Vision loss, central}
    \label{fig:bild19}
\end{subfigure}
\hfill
\begin{subfigure}[t]{0.24\textwidth}
    \includegraphics[width=\linewidth]{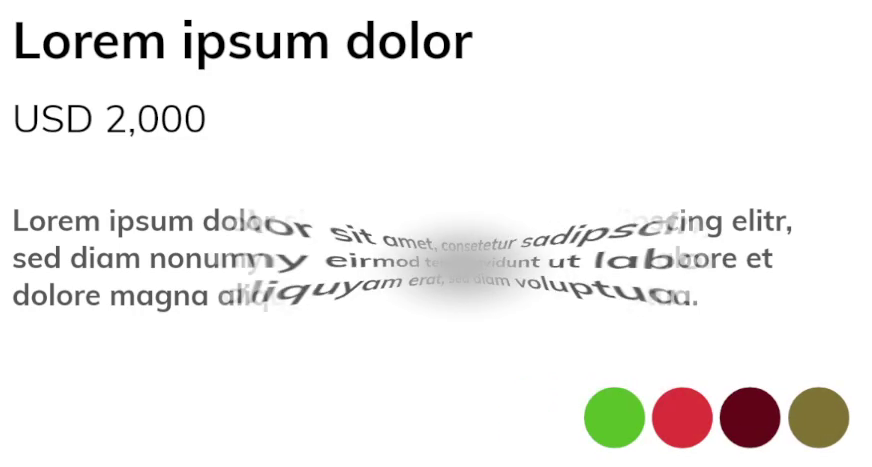}
    \caption{Distortion}
    \label{fig:bild20}
\end{subfigure}
\Description{A 4-by-5 grid showing simulated visual impairments. Each cell displays a different effect, including blurriness, smudging, color loss, and image distortion, to illustrate how simulations of visual conditions can affect perception.}
\caption{Overview of the shaders available in \VIPSIM. It is important to note that each shader still has differences in parameters and severity. Nystagmus is not represented here as the motion is not depictable.}
\label{fig:gridview}
\end{figure*}

\VIPSIM builds on work by \citet{jones2020Seeing}. We describe our additional implementation in the following. An overview of the applied shaders can be found in \autoref{table:shaders} and \autoref{fig:gridview}. % will now describe the shaders we provide and evaluate our contribution. Note that we build our codebase by adjusting and extending the codebase of \citet{jones2020Seeing}.

\begin{table*}[ht]
\centering
\footnotesize
\begin{tabular}{>{\RaggedLeft\arraybackslash}p{0.15\linewidth}p{0.08\linewidth}>{\RaggedRight\arraybackslash}p{0.65\linewidth}}
\toprule
 Name & Eye Tracking (required) & Parameters\\
\midrule
Vision loss central & yes & Size (from 0 to full-screen size) \\
Hyperopia & no & Visual acuity in cycles per degree (CPD) 0.01 (strong effect) to 30 (minor effect)\\
Color vision deficiency & no & Type (protanomaly, deuteranomaly, tritanomaly, and monochrome), Severity (0 to 100 percent) \\
Contrast Sensitivity & no & Brightness (-1 to 1), Contrast (-1 to 1), Gamma (0 to 1)\\
Metamorphosia pointwise & yes & -  \\ 
Nystagmus & no & Speed (0 to 1 time in seconds to rise to the amplitude), Amplitude (0 to 20 percent of screen width) \\
Retinopathy / Floaters & yes & Color (black or white), Opacity (0 to 1), Density (0 to 2500 dots), Speed (0 no movement to 1 fast movement), Centering (in a circle), Circle Radius (from 0 to full-screen size)\\
Teichopsia & yes & Strength (0 to 1 opacity and alliance of scintillating)\\ 
Metamorphosia overlay & no & Speed (0 static to 1 strong displacement), Frequency (0 to 1), Amplitude (0 to 1)\\
Glare vision / photophobia & no & Intensity (0 to 1), Blur (0 to 1), Threshold (0 to 1)\\ 
Vision loss, peripheral & yes & Size (from 0 to full-screen size)\\
Cataracts & no & Severity (0 to 1), Frosting\\
In-Filling & yes & Size (from 0 to a quarter of full-screen size)\\
Double Vision & no & Displacement (from 0 to a quarter of full-screen size)\\
Distortion & yes & Radius (from 0 to full-screen size), Suction strength (0 to 1; the higher, the narrower), Inner radius (from 0 to outer radius), Noise (0 to 1 displacement during suction)\\
Foveal Darkness & yes & Size (from 0 to full-screen size), Fade (0 no to 1 fade starts at the center), Opacity (0 to 1)\\
Flickering Stars & no & Radius (from 0 to full-screen size), Fade\\
Detail Loss & no & Severity (image is clustered into 10 to 1000 clusters)\\
\bottomrule
\end{tabular}
\caption{The symptom shader \VIPSIM incoperates to simulate impairments. Some shaders include eye tracking due to the requirement for gaze contingency of the symptoms. Impairments like a CDV simulation do not require eye tracking.}
\Description{The table lists various visual impairment simulation types, whether eye tracking is required for each, and their adjustable parameters. There are three columns: Name (of the visual effect), Eye Tracking (required) (yes or no), and Parameters (technical settings used to simulate each condition).}
\label{table:shaders}
\end{table*}

\paragraph{Field Loss}
\label{sec:cvs}
Field loss describes a condition where parts of the field of view are impaired. For example, they can be completely blurred or not perceivable. Peripheral field loss, central field loss, or individual spots are affected in the visual field~\cite{wang2024How}. AMD is a prominent disease that affects the central field. Taylor et al. gathered descriptions of AMD patient's words and phrases that were used to describe where: Blur ("Not Clear", "Out of focus", "Fuzzy"), Distorted ("Bendy", "Crooked", "Wavy"), Missing parts ("Patchy", "Black parts", "Grey parts", "Words dropping from page") \cite[p. 4]{taylor2018Seeing}. For VIPs with metamorphopsia, instead of seeing lines straight, a somewhat wavy or distorted perception in an area could occur~\cite{nei_macular_degeneration}. Glaucoma, through damage to the optic nerve, can also lead to field loss in all areas~\cite{kingman2004glaucoma}. Furthermore, retinitis pigmentosa (RP) was a degeneration of rod and cone photoreceptor cells in the retina, narrowing the peripheral field to tunnel vision (peripheral field loss). Central Field loss can also occur since RP can vary widely among individuals~\cite{hartong2006Retinitis}. Diabetic retinopathy is characterized by damage to the retinal blood vessels, which can lead to vision impairment and blindness. Patients may experience blurred vision, floaters, and dark spots, significantly impacting their visual acuity~\cite{fong2004diabetic}.
We improved the Central Vision Loss simulation by not just occluding the central visual field with a mono-colored shape, but also the loss of vision by reducing the render resolution~\cite{taylor2018Seeing, jones2020Seeing}. An elliptical shape with a gray gradient from the center to the outside was used to map the LOD (Level of Detail), which allows for a sample rate on the rendered image. Starting from the middle, where only light, dark, and colors are visible, the visualization has a gradient from being extremely blurry (Low LOD) in the center to being sharper based on the distance to the center (high LOD).
The scale of the applied overlay (from 0 to full-screen size) is adjustable.
This shader uses the eye-tracking input to displace the central point of vision based on the point of gaze.

\paragraph{Hyperopia}
\label{sec:hyper}
Refractive errors, for example, hyperopia, often result in blurred vision~\cite{nei_refractive_errors}; therefore, we applied a Gaussian kernel matrix to the screen. The CPD, a visual logarithmic visual acuity measurement, can be set (range: 0.01 (strong effect) to 30 (minor effect)). 

%\begin{equation}
%\begin{split}
%\sigma &= \frac{1}{2 \pi \cdot \text{CPD}} \\\\
%\sigma &= \sqrt{2 \ln(2)} \cdot \sigma \quad  \\\\

%\sigma &= \sigma \cdot \text{pixel\_per\_degree} \quad \\\\

%\text{kernelSigma} &= \sigma
%\end{split}
%\label{eq:kernel}
%\end{equation}

\paragraph{Color vision deficiency}
\label{sec:cvsshader}
CVD makes specific wavelengths of light not perceivable. There are three types of CVD: deuteranopia or red-green blindness, where L-cones are missing; Tritanopia, where S-cones (blue wavelengths) are missing; and Protanopia, where red and green cones send the same signal. Also, with achromatopsia or monochromatic vision, no cones are available at all, which leads to complete color blindness~\cite{wang2024How}.
We took the fixed CDV calculation matrices of the impairments: protanomaly, deuteranomaly, tritanomaly, and monochrome of related work~\cite{jones2018Degraded, machado2009Physiologicallybased}. These filters then recolored the underlying screen accordingly and modified (removed) the colors for each CDV. A parameter for the severity (0, no appliance to 100, fully appliance of the matrices) is implemented. We relied on the correctness regarding the realism, as a lack of color vision means the absence can not be evaluated.

\paragraph{Contrast Sensivity}
\label{sec:cs}
Contrast sensitivity (CS) refers to the capacity to discern fine details and sharp contours of small objects and to detect subtle variations in shading and patterns. 
Several factors can impact CS. In the early stages of cataracts (clouding of the lens), CS can decrease, particularly for low spatial frequencies, independent of visual acuity. Diseases affecting the visual pathways, including glaucoma and AMD, can significantly diminish CS~\cite{kaur2024Contrast, stalin2020Relationship}.
Our shader enables adjustments to brightness (-1 to 1), contrast (-1 to 1), and gamma (0 to 1), similar to those in image editing software. This combination reduces contrast and adds a grayish overlay.

\paragraph{Metamorphosia}
\label{sec:meta}
It displaces the pixels along two lines on the screen. Initially, it was created to mimic the visual distortion of AMD, in which patients reported perceived curves instead of straight lines. However, our participants with AMD disapproved of the visual effect. The second metamorphosia shader distorts the whole screen by applying a wavy displacement.

\paragraph{Nystagmus}
\label{sec:nystag}
This shader animates the screen in a jumping manner to mimic the random eye movement of nystagmus patients. In this symptom, the screen is continuously shifted to the right in a jerking motion. The jerking motion can be configured via the amplitude and speed. 

\paragraph{Retinopathy}
\label{sec:retino}
This shader spawns dots at random positions. These are shaped based on a Perlin noise pattern~\cite{perlin1985Image} in a pre-generated texture on the screen. These dots have a gradient and can be set to black or white. One can also set the intensity, density, floater size, and speed. Also, the option to center the occurrences of the dots on the point of gaze was added.

\paragraph{Teichopsia}
\label{sec:teich}
An overlay with a star form is applied to the underlying screen, augmenting the colors below while adding a scintillation of colors.

\paragraph{Glare Vision / Light sensitivity / Photophobia}
\label{sec:lightsensivity}
Light sensitivity (photophobia or glare sensitivity) leads to an unpleasant experience when exposed to bright light sources. Also, a blooming and blurring effect occurs~\cite{wang2024How}.
This shader generates a bloom effect that makes bright areas of an image appear to glow. It achieves this by blending the original screen with an additional texture that enhances the luminosity and applying blurring to soften the glow. As a result, the shader produces a visual effect that emphasizes bright regions in the screen, giving it a smoother, more radiant appearance.

\paragraph{Vision loss, peripheral}
\label{sec:vlp}
This vision loss shader is an inverted version of the central vision loss. Therefore, it shows a kind of tunnel vision while still allowing one to perceive light, darkness, and some colors outside the tunnel. 

\paragraph{Cataract}
\label{sec:cata}
We implemented a two-part shader for the cataract shader. First, we removed the screen's contrast and brightness to simulate the lens's clouding. Second, we implemented a frosting shader in which pixels are displaced with a Perlin noise pattern.

\paragraph{In-Filling}
\label{sec:inf}
For this effect, the pixel values surrounding the 3 pixels (at a configurable fixed position) are copied or relocated for each pixel. This generates a visual effect that includes a point, and elements on the screen can disappear.

\paragraph{Double Vision}
\label{sec:dv}
For the double vision shader, the original screen is copied and translated with an intensity to the left and the right. Both screens contribute 50 percent opacity.

\paragraph{Distortion}
\label{sec:distr}
Two things happen in the distortion shader: First, a vortex effect is applied to the screen, and second, a gray shape occludes the screen. For the vortex effect, pixels are drawn to the center point. The shader allows the strength of the suction, radius, and noise applied to the original pixels while they are drawn inwards to be set.

\paragraph{Vision loss, Darkness}
\label{sec:vld}
A simple occlusion of the screen is achieved by applying a gray dot on the screen.

\paragraph{Flickering Stars}
\label{sec:flick}
This shader randomly spawns points on the screen where the LOD, therefore the level of detail, is reduced in the affected area.

\paragraph{Detail Loss}
\label{sec:detailloss}
The detail loss shader is a different approach to making the screen blurry than the hyperopia. A pixelation shader (\textit{Detail Loss}) is applied to reduce the detail amount. This is achieved by looking at the pixel coordinates with a calculated higher pixel size. To make the effect more convincing, we decided to interpolate between the clusters to retain a mosaic-like appearance.

\subsection{User Interface}
\begin{figure}[ht]
    \centering
    \includegraphics[width=0.5\textwidth]{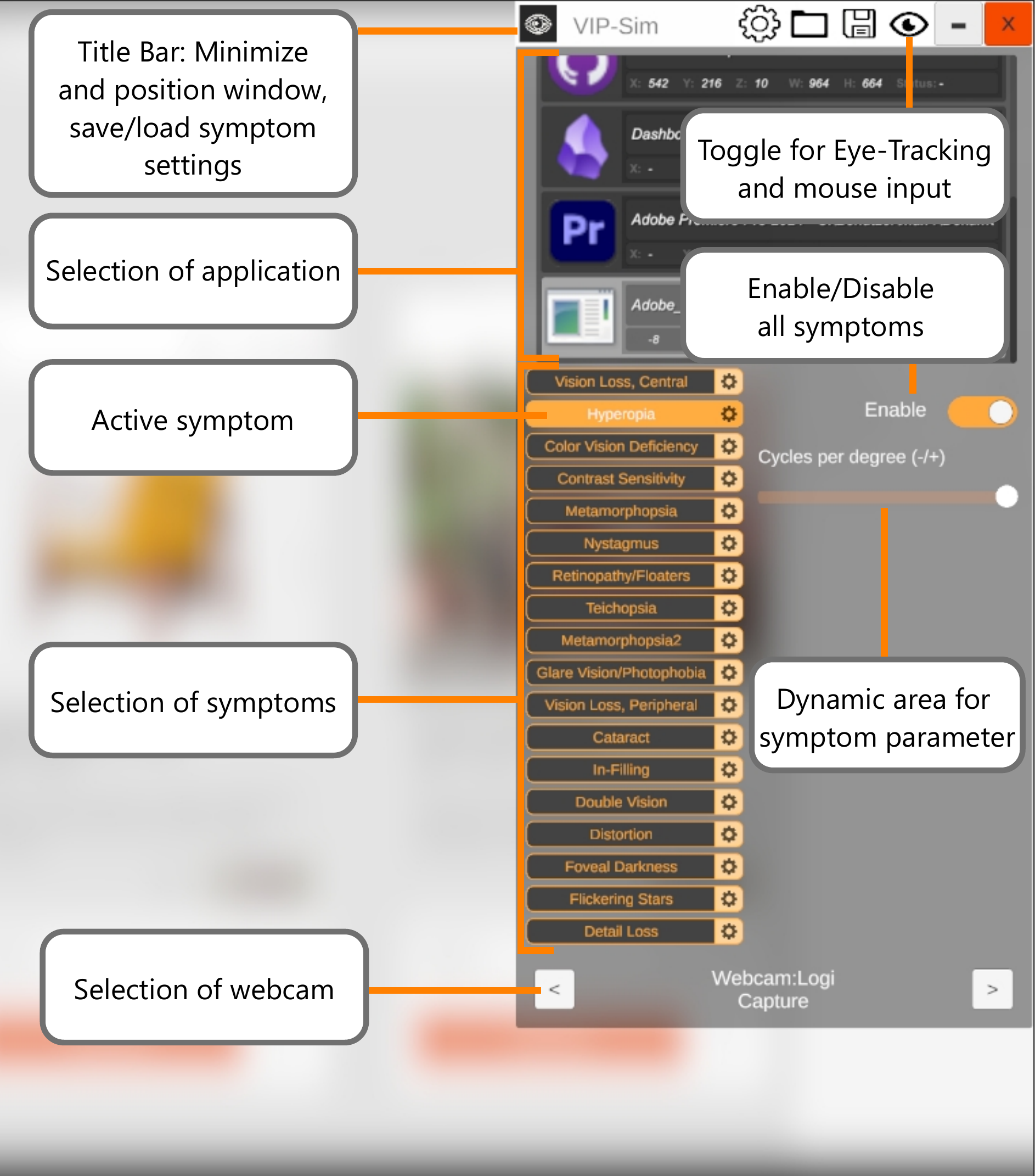}
    \caption{\VIPSIM overlays other applications (e.g., Adobe XD) without interrupting their functionality. The upper section of the UI allows users to select the application for integration. In the lower section, symptoms can be toggled and their associated parameter settings adjusted. An "Enable" toggle allows activation or deactivation of all symptoms simultaneously.}
    \label{fig:ui}
    \Description{This figure depicts the UI of VIP-Sim, which is overlaid on another application. The elements of the UI are labeled as follows: "Title Bar", for minimizing and positioning the window, "Toggle for Eye-Tracking and Mouse Input", "Selection of Application" (pointing to a list of applications displayed at the top), "Enable/Disable All Symptoms", "Active Symptom", "Selection of Symptoms", "Dynamic Area for Symptom Parameters", and "Selection of Webcam".}
\end{figure}

The UI (see \autoref{fig:ui}) is floating over the underlying program. The application on which the simulation is applied can be selected in the top section. This is made possible with UWindowCapture\footnote{\url{https://github.com/hecomi/uWindowCapture}; Accessed: 22.03.2025} on Windows and mcDesktopCapture on macOS\footnote{\url{https://github.com/fuziki/mcDesktopCapture}; Accessed: 30.01.2025}. The buttons on the left allow the selection of one or multiple symptoms. %Multiple symptoms can be applied at the same time. 
The parameters can be set via the settings wheel.
\VIPSIM can be turned off via a simple toggle.
%Using the settings wheel, the impairment parameters, e.g., their severity or variant, can be set. Using the enable toggle, the simulation is disabled, and the overplay projection is turned off. 
In the bottom section, users can select the input camera. In the title bar, the settings field below can be dismissed so that only the title bar remains visible. The eye symbol \frontaleye on the right can be toggled to switch from the eye tracking to the mouse input mode. This feature was added because some participants showed symptoms of nystagmus (random eye movement) or strabismus (being crossed-eyed), which would significantly worsen the performance of the eye tracker to a point where it is not feasible. 

\subsection{Interview Analysis}
The interviews were locally transcribed using Whisper~\cite{radford2023Robust}. Then, the first author refined the produced transcription and fixed errors, assigned speakers, added annotations regarding what the participant saw at the time, and anonymized the interview if needed. 
%We used the Straussian grounding theory (GT) by \citet{strauss2003Basics}. 
The first and second authors applied open coding for the interview. Axial coding was used to identify relationships between them, and lastly, selective coding was applied to summarize the major categories.
The first and second authors read the transcript independently and applied codes. 
Then, a discussion took place. Conflicts were resolved via discussion among the researchers.

\section{Results and Discussion}
This section presents the results and a discussion of our interview process. We first report whether participants felt that \VIPSIM could replicate their symptoms. Subsequently, the emerging themes regarding the acceptance of and concerns about \VIPSIM among VIPs are addressed. We conclude by introducing an ethical dilemma that emerged during the development phase regarding reducing possible adjustment options for symptoms, as a reduction will inevitably lead to exclusion. As our results are mainly qualitative and thus also are based on our understanding, we opt for an immediate discussion instead of separating them. 

\subsection{Interface Challenges and Coping Strategies}
Participants reported facing challenges with modern UIs daily, particularly when reading text on screens. The problems mentioned were too small or blurred fonts and a lack of contrast between text and background, making reading difficult. Although numerous accessible UIs are available today, there are no universal solutions, meaning these problems occur repeatedly:

\begin{quote}
\textit{''And that's often the case, it's often a patchwork quilt [...] if you say there are accessible forms, there are some, but unfortunately, not all of them are accessible.''} (P6)
\end{quote}

These results echo previous works~\cite{laamanen2024Does, webaim2025million}, showcasing that despite efforts to increase accessibility, there are still challenges to overcome. 

It also showed how users deal with inaccessible UIs. The screen size or distance from the screen is often adjusted to make reading easier. Six participants also use assistive technologies such as special smartphone apps or read-aloud functions. However, these technologies have to be learned beforehand, which can be an additional hurdle:

\begin{quote}
\textit{"Voice-over, that's great. [...] I can't use this voice-over like that, but I have to learn that first. I don't know how to do it."} (P4)    
\end{quote}

This showcases that, despite tools for designers exist~\cite{manca2023transparency, webaim2025million}, these are either not used or are insufficient.

\subsection{Resemblance, Acceptance, and Considerations of \VIPSIM}\label{sec:accuracy}

\subsubsection{Resemblance}
We also asked participants about the resemblance of shaders, which are overviewed in \autoref{table:shader_pipeline_map}.

\begin{table}[ht]
\centering
\footnotesize
\begin{tabular}{|p{0.28\linewidth}|c|c|c|c|c|c|c|}
\hline
\textbf{Shader Name} & \textbf{P1} & \textbf{P2} & \textbf{P3} & \textbf{P4} & \textbf{P5} & \textbf{P6} & \textbf{P7} \\
\hline
Vision loss central &  - & X &  X &  X &   & X &   \\
Hyperopia & X &  X & X &  X &   & X &   \\
Color vision deficiency & - & X &  &   &   & X &   \\
Contrast Sensitivity & X  &  X &  X &  X &   & X &   \\
Metamorphosia point & - &   &   &  ! &   & ! &   \\
Nystagmus &   & o &  & ! & o &   &   \\
Retinopathy / Floaters &  &   &   & ! &   & ! & -/X \\
Teichopsia &  &  &   & ! &   & -/! &   \\
Metamorphosia overlay &  &  &   &  -/!  &   &  ! &   \\
Glare vision & - &  X & X &   &   &  X &   \\
Vision loss, peripheral &  &  &  X &   &   &   &   \\
Cataracts &   & - &   & - &   &  X  &   \\
In-Filling & - &  &   &   &   &  ! &   \\
Double Vision &   & X &  X &    &  &  X &   \\
Distortion & * &   &   &   &   &  ! &   \\
Foveal Darkness & * & X &   &  X &   & X &   \\
Flickering Stars &   &  &   & *  &   &  ! &   \\
Detail Loss &   &   &  &   &  * &  X &   \\
\hline
\end{tabular}
\caption{Assignment of symptom simulation shaders to participants (P1–P7).
A checkmark (X) indicates the shader was appropriate without changes. A dash (–) means differences were noted and the shader was adjusted afterward. An asterisk (*) shows the shader was added after this participant. A circle (o) means the shader was inappropriate. An exclamation mark (!) indicates the participant found the shader accurate, but their evaluation was based on memory due to temporary symptoms (e.g., high blood pressure) that had resolved or worsened.}
\label{table:shader_pipeline_map}
\Description{This table summarizes how visual symptom simulation shaders were assigned and evaluated by participants P1 through P7. Each row lists a specific shader representing a visual impairment symptom, and each column (P1 to P7) records the response of a participant.
Symbols used in the table: X: The shader was appropriate without changes. -: Differences were noted, and the shader was adjusted afterward.*: The shader was added after this participant. o: The shader was inappropriate for this participant. !: The shader was found accurate, but the evaluation was based on memory due to temporary symptoms (e.g., high blood pressure) that had since changed.}
\end{table}

Five participants stated that \VIPSIM effectively simulated their symptoms. Although the medical conditions of our participants differed, the central vision loss shader was especially useful, was well received, and covered the same symptoms for different impairments (P2, P3, P4, P6): 

\begin{quote}
\textit{"So I see it exactly the same way, I would say. Maybe a little worse sometimes.} Okay. So, worse in the sense of bigger then? \textit{Yes."} (P3)
\end{quote}
The LOD and the degree at the edges also reflected their experiences. When asked, "Are there still differences somewhere?", (P6) stated:
\begin{quote}
\textit{No, that's fine. Even with the flowing out, that's good.}"
\end{quote}

Also, participants perceived the hyperopia, or blur shader, as representative of their symptoms. (P2), (P3), and (P6) could adjust the severity themselves and stated that the shader successfully portrayed their perception. We noticed that this shader seemed appropriate for a general portrayal of visual acuity loss. However, there is no one-size-fits-all solution. Often, instead of blurring, the loss of detail is a factor~\cite{goodman-deane2007Equipping}. For (P5), the shader did not represent the perception as expected, even though the initial description suggested it would: 
\begin{quote}
\textit{"No, I don't have any blurring. [...] but I just recognize more details when I get closer [...]"} (P5)
\end{quote}

Therefore, we asked how we could visualize this effect and found that a pixelation shader was most appropriate. (P6) also indicated that the pixelation shader (Detail Loss \autoref{fig:gridview} (g) in \VIPSIM) could replicate their symptoms better than the blur shader.

The CVD evaluation was not feasible based on our participants' perceptions because severe deficiencies and other impairments made colors no longer perceivable to them. However, during our study, while changing the severity parameters of the shaders, thereby reducing the intensity of target colors, participants with CVDs did not notice any differences.

We, therefore, argue that, aligning with \citet{machado2009Physiologicallybased}, the CVD shaders might replicate this symptom.

All participants described the CS shaders as closer to their perception than the specially crafted cataract shader. If the CS shader was applied, participants could no longer read the given texts, while it was possible for the supervisor. Also, the CS shader was perceived as resembling the participants' symptoms:

\begin{quote}
\textit{"No, it was nothing else. It was altogether everything darker, grayer, dark gray, actually not black, but dark gray."} (P4), \textit{"That's about right. It's a more washed-out image. The color intensity got in the way. [...] As if there is always a light fog [...]"} (P2).
\end{quote}

Despite being well-received, the retinopathy shader required some changes. (P7) had optic atrophy, but they felt the shader could replicate their symptoms. (P7) reported that their impairment does not lead to blurriness:
\begin{quote}
\textit{"It's just completely blurred and it's not like that for me."} (P7)
but rather 
\textit{"Just tiny little dots. But I can still see everything around me. So I see you now too. And the furniture, the wall."} (P7) 
\end{quote}

However, we adjusted the shader as the parameters could not fully convey their impairment. The dots had to be
\begin{quote}
\textit{"Smaller. Denser. [...] centered in the middle"} (P7) 
\end{quote}

The light sensitivity shader was especially well-received. It could simulate the effects of a sunny day on the environment and the blurriness of bright interfaces:
\begin{quote}
\textit{"The image then starts to get a spongy consistency in strong light, as it is now (points to screen). A spongy consistency. Very bright objects then become dazzling and lumpy."} (P2)
\end{quote}

All our light-sensitive participants (n=6) use special filters or sunglasses to countermeasure the display's brightness and perceive the effect of the shader during the study. Challenges with bright interfaces especially appeared, and all participants described the shader as a close replication of their symptoms.

(P1) reported that they have a visual distortion in their perception: 

\begin{quote}
\textit{"I can see a large area, but can no longer concentrate in the middle, where the picture collapses. [...] I can't describe it like that now... It's like a pull inwards; it all comes together in the middle."}. (P1)
\end{quote}

Despite the in-filling shader’s near success, its output did not align with (P1)'s perception, resulting in an addition of the distortion shader (\autoref{fig:gridview} (f)) to \VIPSIM.

Lastly, the double vision shader was also perceived as effective by (P4), who reported that this effect occurs with high blood pressure:

\begin{quote}
\textit{"That's about right. Yes. It's as real as double vision."} (P4)
\end{quote}

No participant approved our shaders for metamorphosia, teichopsia, and nystagmus. This disapproval is either because none of our participants had these symptoms (teichopsia) or because these visualizations did not resemble their perception (metamorphosia, nystagmus). 

The metamorphosia shaders were designed to resemble the effects of a macular condition. While the metamorphosia shader displays a displacement along two lines, participants described the occurrence of a vortex-shaped displacement effect. In our final version, we added disclaimers to these shaders within the UI.

\subsubsection{Acceptance}

We asked generally about \VIPSIM being used by designers ("Imagine a designer uses a tool to simulate the symptoms of your vision impairment, how are your feelings about that?"). We also asked: "Do you have any concerns regarding designers using simulators while designing?". Five participants appreciated the idea of \VIPSIM being used by designers. 
%We classified the feedback into six major topics: (1) Acceptance of simulation, (2) possible use cases and benefits, (3) expectations of \VIPSIM, (4) comprehensiveness, (5) fear of exclusion, and (6) industry acceptance. 

\begin{quote}
\textit{“It would be beneficial for me. Yes, of course.”} (P4) Do you have any concerns if a designer uses this now? \textit{“No. Why should I have I don't quite understand. If you use something like that, you use it to make it easier for people like me. make it easier”} (P4)
\end{quote}

\begin{quote}
\textit{"Maybe from my case, what you just showed me, you could imagine what a person can see and you could pay attention to it, where are the weak points, where are the strong points. Such an application would help to give the idea for the developers."} (P3)
\end{quote}

However, our study indicated that negative attitudes towards simulators emerged whenever the initial shader failed to represent the impairment (P1, P5, P7), underscoring the importance of precise simulation. Participants, where an immediate, accurate representation of their symptoms was feasible, did not express any concerns. Instead, they highlighted positive aspects of the simulator, such as the belief that it could foster a better understanding of their condition. They stated that it could help identify potential issues early on, preventing inaccessible design. 

This provides nuance to work by \citet{bennett2019Promise} who reported that people often feel misunderstood by simulators, holding negative attitudes towards them. These attitudes are frequently associated with the fear that such tools may reinforce negative images and stereotypes. In contrast, five of our participants reported feeling that \VIPSIM visualizes their impairment (see \autoref{sec:accuracy}) and stated that \VIPSIM effectively helps others better understand their visual experience. This also contrasts with related work by \citet{maher2024Stop}, where previous methods were often perceived as limited in terms of realism, contributing to negative perceptions. Therefore, we believe \VIPSIM could be the first simulator achieving the required realism.

%Participants expressed a desire for accessible designs.

%\begin{quote}
% \textit{"Simple, as simple as possible. Because for us, let's say, this is simply important for our group of people. Or for me in particular. It's no use to me if I have a thousand pieces of information. It's of no use to me. I have to be able to access that, whether it's a computer screen
% or a cell phone or a tablet, I don't care, I have to be able to look at it and say, okay, that's it, that's it. And of course, what is really desirable, a reasonable contrast."} (P5)
%\end{quote}

Four participants (P2, P3, P4, P6) indicated that a simulator like \VIPSIM might support accessible design and have a positive impact on their lives. Regarding underlying ethical questions, they stated that simulation poses minimal risk and its use is not disadvantageous: 

\begin{quote}
\textit{"After all, a simulator is there to technically represent life, to do in your head, what could it look like? They're all just simulation games, simulators. You can't break anything, you can't upset anyone with it. It's just a toy for you alone. Let's put that in quotation marks. It's just a toy for you, who uses it to represent something or to represent some kind of process. So I have zero concerns about that."} (P2)
 \end{quote}

Although (P2) and (P5) have experience in designing software, it is essential to highlight that VIPs may not necessarily have a clear understanding of the design process~\cite{bennett2019Promise}.

Without the simulation, the participants stated, nothing would change either. This is similar to \citet{tigwell2021Nuanced}, where 13 of 17 participants argued similarly. Nonetheless, the 13 participants who were positive still had some reservations. 
%However, participants also offered other perspectives similar to \citet{tigwell2021Nuanced}. 
This aligns with (P1), who confirmed that the simulation replicated their symptoms, but they still wished to be personally involved in developing accessible designs: 

\begin{quote}
\textit{"So what we've done now can't be done any better. I think it was optimal. To make what happens to my eye more or less explainable. With all the effort you put in and make the technology right and good, everything, but the human eye or what I have inside me simply can't one hundred percent."} (P1)
\end{quote}

Specifically, the completeness of the simulation was a concern. Participants reported that while the simulation worked very well for them, they knew others in the community who experienced different effects, highlighting the issue of the diversity of visual impairment. Especially mentioned by our participants are visual hallucinations apparent in e.g., Charles Bonnet syndrome~\cite{jan2012visual}. %Haben such nichts für das syndrome entsprechend teilnehmer für die evaluation zu finden wird sehr schwer

\begin{quote}
\textit{"You're not the center of the world yourself, you have a very individual illness, I say. In other words, your medical history. And when I'm here with my visually impaired group, everyone sees differently. Everyone's vision is bad, but everyone's vision is different."} (P6)     
\end{quote}

\subsubsection{Considerations}
 
Concerns were raised about whether the simulation would provide added value, given that impairments might not be compatible. 

\begin{quote}
\textit{"That may be the case in my situation I see it that way, but you should also consider other illnesses. Some people see differently. The visual impairment is not always the same for all people. It depends on the disease."} (P3)
\end{quote}

One participant (P5), who felt that \VIPSIM could not replicate their impairment, expressed the view that technical feasibility can never be fully achieved and that the only proper solution is direct contact with VIPs. Additionally, there was dissatisfaction with standardized norms and guidelines.

\begin{quote}
\textit{"I always say, ask the people. [...] Everyone has a different eye disease. And everyone tells you something different. There is no one solution for everyone. There is no such thing. [...] I always say, ask the people. Let them try it out. And then they say, that sucks and that's good. [...] And as I said, for me there is no one solution. There is no such thing.[...] Then it's always, yes, God, that corresponds to the DIN standard blah blah blah and blah blah blah. So if I had to work with it for eight hours, I'd be banging this stuff around your ears."} (P5). 
\end{quote}

This aligns with work by \citet{kim2018EmpathD} and \citet{power2012Guidelines}, who strongly contradict the notion that financial constraints should be a reason for exclusion. 

Another participant reported that involving VIPs in the process is challenging because they frequently experience frustration and a lack of commitment to addressing their impairments:

\begin{quote}
 \textit{"But most visually impaired people who say, “I? I know I'm healthy, I don't have anything." There are a lot of people who don't want to have anything to do with us. As I said, the idea is great[...]. Or I just know how stubborn our people are. [...]  So I always have the feeling we exist under the table, quite a lot, but quite publicly, people somehow never want to admit that they actually belong"} (P2)
\end{quote}

VIPs see the application of \VIPSIM in public spaces 
\begin{quote}
\textit{"When introducing anything that practically gets into the public domain."} (P6)
\end{quote}
, in raising public awareness, 
\begin{quote}
\textit{"Does he now want to show the public how a visually impaired person perceives the world?"} (P2),
\end{quote}
and in the design of products and creating awareness 
\begin{quote}
\textit{"I can't get away from the track, that the one can then empathize with us or can somehow adjust to our view of the world."} (P2).
\end{quote}
However, a recurring theme was the acceptance of the software, especially among designers. Concerns were offered by (P6) that there is no enforcement for using simulators. Additionally, concerns were raised that while a simulator can identify problems, there is no enforcement to solve them, and it is precisely the solutions that are most relevant for VIPs. Another point mentioned was the simulator's effectiveness. While \VIPSIM draws attention to some partial issues, it does not make essential aspects such as the readability for Voice-Over or the correct structure of websites the WCAG~\cite{wcag2024} suggests apparent. 

\begin{quote}
\textit{"So for the developer, I'd say he has a bit of an idea of what possibilities, what needs to be taken into account, I'd say. Where are the fields anyway? Because the person with normal vision doesn't even know that. [...] No, he can't actually do anything to the disadvantage. So if he knows the things, the most he can say is that I've recognized the problem, but it's too time-consuming for me to fix it [...] But I mean, that's another question, how to simply make something like that the standard, let's put it that way. [...] So now, for example, just when I can open these PDFs, there's nothing in here, for example. So yes, I could open a PDF now, for example. Yes, you can open it, but you can't test whether it's accessible. Whether it reads aloud well, for example"} (P6)
\end{quote}

An additional concern raised, particularly by participants with degenerative conditions, is that it may lead to designs that help them now but lose their positive impact in the future, as they might lose their remaining vision.

\subsection{The \textit{MonoWitness protocol}: Valuable Insights and Identified Gaps}
Our process helped us understand and replicate visual impairments. The CVD, CS, and Amsler grid tests allowed us to identify the impairments quickly and efficiently.

In the first part of the test (see \autoref{fig:meth} A), we could determine the impact of their field loss for each participant. We were concerned that showing simulated impairments to VIPs would be difficult, as participants would experience the impairments through the simulation and due to their own \textbf{actual} impairments. However, this was not an issue, as we surpassed this challenge by letting VIP scan the image, using screen magnification, or having participants with one capable eye. Additionally, showing and asking participants, as done by \citet{taylor2018Seeing}, worked very well for us, and participants (P2, P6) could adjust the simulator themselves.

Another useful advantage in future projects is using one eye at a time and then both eyes. As \citet{krosl2020CatARact} demonstrated, where participants had only one successfully treated eye, we could also identify different symptoms using one eye and both eyes.

We also noticed that identifying CVD and CS was particularly challenging, as there is no internal compensation for these conditions. This means that a person with CVD cannot judge whether the presented visuals are accurate since they do not perceive any change. A person with contrast sensitivity issues might not experience the perception created by the shaders as intended. In our process, we evaluated the CS shaders by having participants read text with less contrast at varying distances and then checking whether even fewer lines were readable with the shader activated. While this method helps identify problems caused by low contrast, the shader remains only partially evaluable.

It was also striking that CVD and CS often appeared as accompanying symptoms~\cite{krosl2020CatARact}. Even when participants initially denied experiencing these issues during the initial assessment, our tests indicated they must be present.

%The use of eye tracking was also crucial, particularly for participants affected by central field loss, where this feature became the decisive factor in understanding their visual impairment.

Ultimately, our process worked very well for participants who had impairment in only one eye or could explicitly recall a time without impairment. However, assessment was more difficult for participants who had a congenital impairment since birth, as they lacked a reference point to compare their perceptions to others. 

\begin{quote}
\textit{"I'm not used to it any differently from the other birth. I can't imagine what it feels like for you now. That is always the difficulty that we, let's say now, have with the congenitally blind or congenitally visually impaired. I don't know any other way. For me, when I look out of the window here now, it's completely normal for me. Because I don't know any other way."} (P6).
\end{quote}

In this case, the \textit{MonoWitness protocol} also reached its limits.

\subsection{The Ethical Dilemma of Comprehensive Impairment Coverage}

To introduce this subsection, we would like to highlight a point that both we, the authors, and the participants observed:

\begin{quote}
\textit{"That you say you have a coverage of 90 percent plus, let's put it this way, of the visual impairments. So you can't go down to the last specification. That's always the case. As I said, you can't expect the developer to do that either. So what I've noticed is that I think you have a huge task, you'll probably have to limit yourself to one of these effects somewhere later on."} (P6).
\end{quote}

\VIPSIM offers a vast range of settings. With 21 symptoms that can be activated simultaneously, $2^{21}$ possible combinations occur. Additionally, each symptom has parameters, further increasing the number of potential settings. The goal of this complexity is to achieve a high degree of symptom coverage and ensure that \VIPSIM can accommodate the impairments of as many VIPs as possible. However, the number of parameters and functionalities eventually reaches a point where they become impractical to apply effectively. One consequence would be to reduce the number of parameters by fixing them to values that represent the average user (and thereby the number of symptoms). This, however, inevitably leads to exclusion. Also, interaction effects of impairments that are possible to visualize with \VIPSIM should not be neglected. \citet{andrew2024light} reports a case where a designer mentioned an interaction effect between having dark mode (which is generally considered more accessible by designers) and astigmatism. The other designers did not mention such issues~\cite{andrew2024light}. \VIPSIM could help designers to understand how impairments affect each other.

An alternative approach could be prioritizing symptomologies based on their prevalence, for example, based on data from the WHO. However, we do not recommend this, as it would again disadvantage a minority of individuals, particularly those whose impairments significantly impact their lives. Should we, therefore, focus only on impairments that "significantly impact their lives"? We strongly argue against this.

The question remains: where should the focus lie? While we cannot provide a definitive answer, we observed that simulating symptoms instead of impairments leads to accompanying various impairments. Maybe this could be a future direction. For example, our process revealed that shaders designed to simulate specific impairments, such as central field loss, could also be applied to other impairments as accompanying symptoms.

One existing solution involves using personas, which come with their own challenges, such as offering a very narrow perspective and potentially leading to stigmatization~\cite{bennett2019Promise}. From our sessions with participants, we also gathered various simulator configurations. To counteract the adverse effects of personas, we suggest a cycling approach. In this method, the simulator would interpolate between impairments in a random sequence, allowing one to perceive a broader range without settling on a specific configuration by manually setting them.

\subsection{Open Challenges: Balancing Realism, Technical Limitations, and Coping Strategies}
While most of our proposed shaders (20 of 24; 83.33\%) resembled participants' perceptions, we acknowledge several ongoing challenges. First, the human eye and brain possess internal mechanisms for compensating for visual defects. Some participants reported perceiving colors even though they are colorblind or experiencing visual infill based on what they expect to be present. These phenomena are difficult to replicate in a simulation because the simulated information does not exist in the source. Similarly, for participants with nystagmus, many reported compensating for their eye movements and thus not perceiving the jerking motion of their eyes at all.

\begin{quote}
\textit{"It would look like this to my counterpart. But I don't notice it at all. I compensate for it."} (P2) \\

\textit{"It just expresses itself like this, for example, when I'm talking to you now, it's difficult for me to always fixate on you exactly."} P(5)
\end{quote}

We asked: "For example, if you look there now, does the blind spot jump for you?"

\begin{quote}
    \textit{Yes, that's also the case. If I look somewhere, the spots I see also jumps. (P3)}
\end{quote}

We then activated the eyetracker.

\begin{quote}
\textit{[...] When I didn't see that just now, I don't think I had seen that. (P3)}
\end{quote}

Because the simulated field loss was always displayed where you can't see anything?

\begin{quote}
\textit{Yes, exactly, because it was exactly in the blind spot, I think. (P3)}
\end{quote}

We encountered technical limitations. The eye-tracking features were particularly challenging to evaluate, as the central point of vision where the field loss occurs shifts with the gaze. This speaks for the quality of the feature, because both simulated and perceived field loss align. However, participants cannot fully approve what they can not perceive.
Finally, we must address the issue of perceived reality. Since participants could not perceive specific parts of our test images, likely, some of our visualizations were also not fully perceivable during the study, which could undermine the results. It is essential to judge our results with care, as our participants found the shaders to be generally applicable. However, they reflect subjective approximations, which is also evident from the non-statistical nature of the data. However, detailed A/B testing of different versions of our shaders or a human-in-the-loop optimization approach to the MonoWitness protocol could further solidify our results. 

\subsection{Limitations and Future Work}
First, our sample size for evaluating visual impairments was relatively small, which made it particularly challenging to account for all possible impairments. We also acknowledge that additional iterations of the UCD process could provide a more comprehensive and well-founded assessment, potentially incorporating additional shaders. Another limitation concerns the estimation of the distance to the screen. Due to the inaccuracy of the eye-tracker’s distance estimation, we chose to omit this feature. However, allowing users to adopt coping strategies is crucial, such as moving closer to the screen. The webcam-based eye tracker UnitEye~\cite{wagner2024UnitEye} has an accuracy of vertical 2.55cm and horizontal 2.40cm and a precision of vertical 0.53cm and horizontal 0.50cm on a 34.5cmx19.5cm screen. Although the Kalman Filter~\cite{welch1997Introduction} was used to smooth out errors, inaccuracies during tracking remain. Our interviews, however, indicate that the eyetracker is precise enough to convey the gaze behavior of our participants. Future work could instead of adhering to the point of gaze at the mouse pointer, project it to regions of interest, or play a moving animation. While these methods can be considered unrealistic, at least they communicate that symptoms like field loss are not static. Furthermore, since the participants were involved in the development and could use the simulator, the halo effect~\cite{nisbett1977Halo} might have occurred. Finally, although two authors conducted the qualitative analysis independently, we cannot entirely rule out the possibility of researcher bias~\cite{chenail2011Interviewing}. 

Future work could investigate the impact of \VIPSIM on designs and the opinions of designers who use \VIPSIM in their daily work. A longitudinal study could be conducted where designers are asked about their design considerations after each use. All the works included in \autoref{sec:rw} focus on short-term studies or one-time usage. However, for such a tool, it is especially interesting how much designers rely on a simulator. Does prolonged usage mitigate or reinforce the adverse negative effects of simulators (see \autoref{sec:promise})? And how are such tools adopted within a workflow? 

Furthermore, it is interesting how \VIPSIM could be used in the education of future designers. Here, pre- and post-tests on their knowledge about accessible designs could be conducted.

\section{Conclusion}
With \VIPSIM, we developed a tool that simulates the sight of people with visual impairments and presented a novel implementation of participatory and user-centered design methods in accessibility.
By incorporating the insights of visually impaired individuals, \VIPSIM demonstrates the power of designing with users rather than simply for them.
In conclusion, \VIPSIM successfully simulates a wide range of visual symptoms (N=21) through a UCD process involving N=7 participants with different vision impairments. Participants felt that we could replicate their symptoms, such as central vision loss, light sensitivity, and blurred vision, to help future designers better understand the challenges faced by VIPs. While the simulator received positive feedback for its accuracy from VIPs, limitations remain regarding its ability to cover all possible impairments comprehensively. Despite these concerns, \VIPSIM has the potential to be a valuable tool for accessible design. However, it should be used alongside direct input from people with visual impairments to ensure inclusive outcomes.

\section*{Open Science}
All source code, installation instructions, and information on required 3rd-party Unity assets for \VIPSIM can be found under \url{https://github.com/Max-Raed/VIP-Sim}.

\begin{acks}
We thank all study participants for their invaluable time, feedback, and cooperation.
This research was funded by the Deutsche Forschungsgemeinschaft
(DFG, German Research Foundation) through the project \textit{“Non-Visual Interfaces to Enable the Accessibility of Highly Automated Vehicles for People with Vision Impairments”} (Project number: 536409562).
\end{acks}

\bibliographystyle{ACM-Reference-Format}
\bibliography{sample-base}

\appendix

\section{Overview of Simulators introduced in Related Work}
\label{apdx:overview}

\begin{small}
\begin{table*}[ht]
    \centering
    \scriptsize
    \caption{Overview of the simulators mentioned in \autoref{sec:rw}. "n.a." indicates that no information is provided.}
    \begin{tabular}{|p{1.5cm}|p{3cm}|p{2.5cm}|p{1cm}|p{1cm}|p{1cm}|l|}
    \toprule
    \hline
        Publication & Symptoms or Impaiments & Visualization & Number of VIPs & Method & Gaze Contingency & Year \\ \hline
        ---Hardware--- & ~ & ~ & ~ & ~ & ~ & ~ \\ \hline
        \cite{zagar2010Low} & glaucoma, cataracts, AMD, diabetic retinopathy, and retinitis pigmentosa. & Nail polish on glasses & 0 & n.a. & no & 2010 \\ \hline
        \cite{goodman-deane2014Simple} & general loss of the ability to see fine detail, & Glasses that occlude the vision & n.a. & n.a. & no & 2013 \\ \hline
        \cite{lowVisionSimulators} & diabetic retinopathy, glaucoma, cataracts, AMD, and hemianopsia & Glasses that occlude the vision & n.a. & n.a. & no & unknown \\ \hline
        \cite{zhang2022Seeing} & field loss, glaucoma, AMD & Glasses occlusion with black color & 0 & n.a. & yes & 2022 \\ \hline
        \cite{almutleb2018Simulation} & central scotoma & Contacts occlusion with black color & 0 & n.a. & yes & 2018 \\ \hline
        \cite{franceswalonker1981Simulating} & central scotoma, Hemianopia, Peripheral field loss & Contacts occlusion with black color & 0 & n.a. & yes & 1981 \\ \hline
        --Desktop-- & ~ & ~ & ~ & ~ & ~ & ~ \\ \hline
        \cite{kamikubo2018Exploring} & Tunnel vision & occlusion with white color & 0 & n.a. & yes & 2018 \\ \hline
        \cite{goodman-deane2007Equipping} & not mentioned & Visual effects displayed on images & 0 & n.a. & no & 2007 \\ \hline
        \cite{giakoumis2014Enabling} & MD, glaucoma, Cataract, Hyperopia, Colour Vision Deficiencies, Night Blindness, Extreme Light Sensitivity, Retinitis pigmentosa  & Visual effects displayed on applications & unknown & n.a. & no & 2014 \\ \hline
        \cite{mankoff2005Evaluatinga} & Blurring, Occlusion, Color modifications & Visual effects displayed on screenshot & 0 & n.a. & no & 2005 \\ \hline
        \cite{schulz2019Frameworka} & cataracts, nyctalopia, CVD, MD, glaucoma & Visual effects displayed on screenshot & 0 & n.a. & no & 2019 \\ \hline
        \cite{krishnan2019Impact} & micro-scotomas & Visual effects displayed on screenshot & 0 & n.a. & unkown & 2019 \\ \hline
        \cite{lane2019Caricaturing} & AMD & Visual effects displayed on screenshot & 0 & n.a. & unkown & 2019 \\ \hline
        --VR/AR-- & ~ & ~ & ~ & ~ & ~ & ~ \\ \hline
        \cite{krosl2020CatARact} & cataracts & Effects applied to 360 video & 5 & Showing visualizations asking for descriptions & yes & 2020 \\ \hline
        \cite{krosl2023Exploring} & refractive errors, AMD, achromatopsia, cornea & Effects applied video–see-through AR & 0 & n.a. & yes & 2023 \\ \hline
        \cite{jones2020Seeing} & glaucoma & Effects applied video–see-through AR & 1 & description of impairment & yes & 2020 \\ \hline
        \cite{jones2018Degraded} & glare, blur, CVD, infilling, distortions & Effects applied video–see-through AR & n.a. & self-reported symptoms & yes & 2018 \\ \hline
        \cite{ates2015Immersive} & MD, diabetic retinopathy, glaucoma, cataracts, color blindness, diplopia & Effects applied video–see-through AR & 0 & n.a. & no & 2015 \\ \hline
        \cite{yao2021Evaluating} & cataracts, diabetic retinopathy, glaucoma  & Effects applied to VR environment & 0 & n.a. & no & 2021 \\ \hline
        \cite{hakkila2018Introducing} & AMD, cataract, myopia, glaucoma & Effects applied to VR environment & 0 & n.a. & no & 2018 \\ \hline
        \cite{zhang2020Developing} & cataracts, glaucoma, MD, color blind,   & Effects applied to VR environment & 0 & n.a. & no & 2020 \\ \hline
        \cite{taewoo2024watchcap} & Tunnel vision & Effects applied to VR environment & 0 & n.a. & yes & 2024 \\ \hline
        \cite{kim2018EmpathD} & cataracts, glaucoma & Effects applied to XR environment & 0 & n.a. & no & 2018 \\ \hline
    \end{tabular}
\end{table*}
\end{small}
%\section{Summary of Visual Impairment Shader}\label{app:}


\end{document}